# A Systematic Literature Review about the impact of Artificial Intelligence on Autonomous Vehicle Safety

A. M. Nascimento[1], L. F. Vismari[1], C. B. S. T. Molina[1], P.S. Cugnasca[1], J.B. Camargo Jr.[1], J.R. de Almeida Jr.[1], R. Inam[2], E. Fersman[2], M. V. Marquezini[3], and A. Y. Hata[3]

***Abstract*—Autonomous Vehicles (AV) are expected to bring considerable benefits to society, such as traffic optimization and accidents reduction. They rely heavily on advances in many Artificial Intelligence (AI) approaches and techniques. However, while some researchers in this field believe AI is the core element to enhance safety, others believe AI imposes new challenges to assure the safety of these new AI-based systems and applications. In this non-convergent context, this paper presents a systematic literature review to paint a clear picture of the state of the art of the literature in AI on AV safety. Based on an initial sample of 4870 retrieved papers, 59 studies were selected as the result of the selection criteria detailed in the paper. The shortlisted studies were then mapped into six categories to answer the proposed research questions. An AV system model was proposed and applied to orient the discussions about the SLR findings. As a main result, we have reinforced our preliminary observation about the necessity of considering a serious safety agenda for the future studies on AI-based AV systems.**

## I. Introduction

Advances in Artificial Intelligence (AI) are one of the key enablers of the Autonomous Vehicles (AVs) development. In fact, AVs rely on AI to interpret the environment, understand its conditions, and make driving-related decisions. Thus, it basically replicates the human driver actions when driving a vehicle. In this context, AI applied to AV has become an important research topic.

AV is a safety-critical system. When operating in an undesirable way, AV can jeopardize human lives or the environment in which it operates. It has the potential to threaten the lives of its own passengers, pedestrians and people in other vehicles, and damage other transportation system elements (e.g. other vehicles and transportation infrastructure). Therefore, it is mandatory to assure AV is safe, mainly when operating on public roads in which resources will be shared with other systems (and people).

Although safety is a mandatory characteristic to AV, and although the researchers seem to agree on the importance of AI applied to autonomous vehicles, they seem to disagree on the AIs impact on AV safety. Many researchers, in special those related to the AI community and AV manufacturers, advocate AI as one of the core elements to enhance AV safety. Their hypothesis is the automation of the driving tasks will lead to a significant reduction of the car accidents. However, other researchers, mainly in the system safety community, argue that AI can potentially jeopardize AVs safety.

This study is the first, as far as we are aware, to map and to organize the related literature and to provide a complete view of the aspects related to both visions, and to subsidize future studies. A preliminary study on the concerns about the differences between AI and system safety mindsets impacting AV safety was published in [1]. In this non-convergent context, this paper presents a systematic literature review (SLR) aiming to present a clear picture of the state of the art of the literature in AI on AVs safety.

This paper is structured into 5 sections. Section II presents details about the research methodology used. Section III presents the data analysis results from the SLR based on the proposed methodology. Section IV proposes an AV system model that is used to orient the discussions about SLR findings. Finally, Section V presents the final remarks.

## II. Research Methodology

This study was performed using the systematic literature review (SLR) method. The reasons supporting the SLR use are: (1) its established tradition as a tool to understand state-of-the-art research in technology-related fields [2]; (2) it helps to understand existing studies and supports readers in identifying new directions in the research field [3]; and (3) it helps to create a foundation for advancing knowledge [4].

The protocol used (Figure 1) was based on the tasks suggested by [5][6] for defining the research questions, identification of search string, source selection, study selection criteria, and data mapping. Also, the protocol followed the recommendations of [7], [8], [4] and [9] for extracting, analyzing, interpreting and reporting the literature-based findings.

This work was supported in part by the Research, Development and innovation Center, Ericsson Telecommunication S.A., Brazil.

[1]Safety Analysis Group from Polytechnic School of University of São Paulo, São Paulo, Brazil (e-mail: alexandremoreiranascimento@gmail.com).

[2]Ericsson Research, Ericsson AB, Stockholm, Sweden (e-mail: rafia.inam@ericsson.com).

[3]Ericsson Telecommunication S.A., Indaiatuba, SP, Brazil (e-mail: maria.marquezini@ericsson.com).

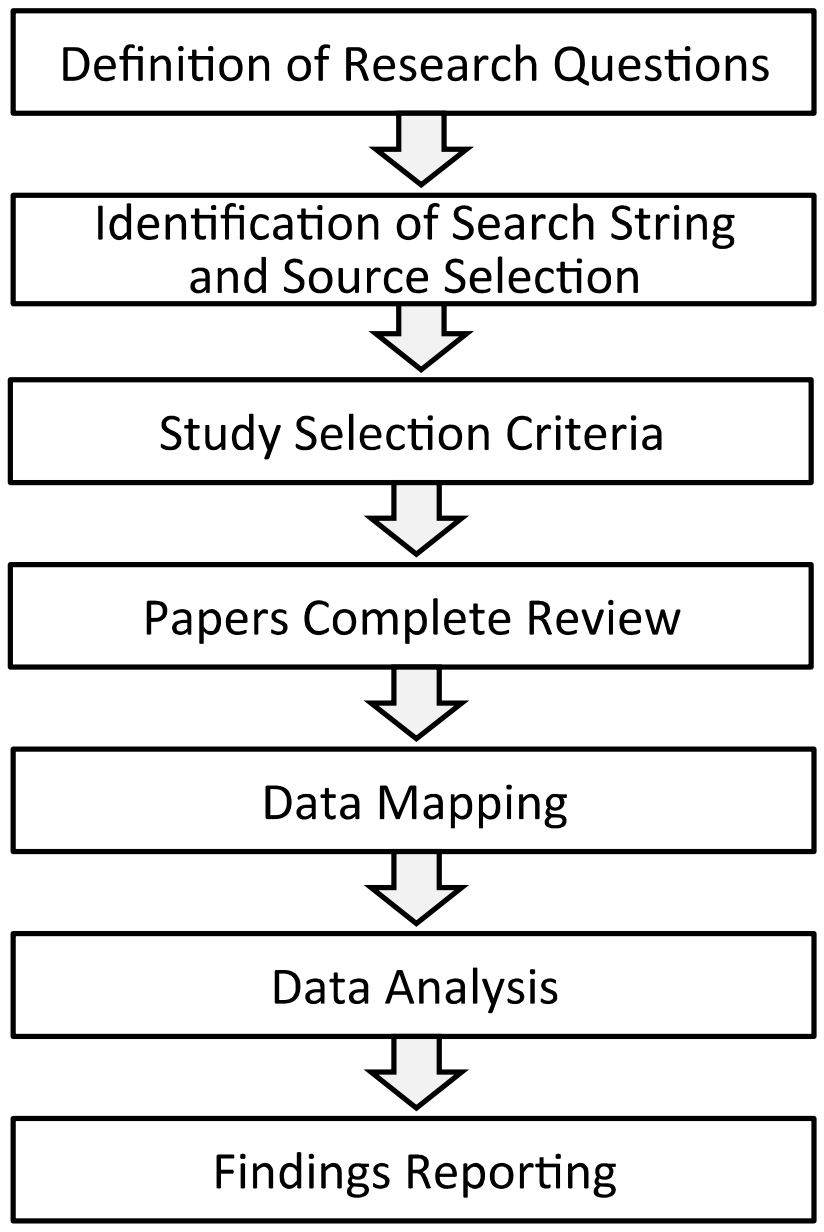

Figure 1. Protocol used to support systematic literature review.

## *A. Definition of Research Questions*

The first step was to define the research questions (RQ). In order to support the research goal of presenting a clear picture of the state of the art in the literature about AI on AV safety, the following research questions were posed:

- RQ1. How do AI-based systems impact system safety?
- RQ2. Which are the topics (context domain) of the studies identified?
- RQ3. Which AI-related techniques are used on the studies?
- RQ4. Which problems do the techniques seek to address?
- RQ5. Which findings are reported by the study's authors?
- RQ6. Which future studies are suggested in these studies?

## *B. Identification of Search String and Source Selection*

The search strategy was structured through the selection of source databases and the appropriate search terms. No date range was used, to ensure that relevant studies were covered, regardless of their publication date. A broad selection of online databases indexing scientific literature was considered: ACM, Engineering Village, ScienceDirect, Scopus, SpringerLink, Wiley and Web of Science (WoS). Please note that IEEExplore is already covered by the selected databases for this SLR study.

The search string was designed based on the synonyms of the 3 main concepts related to the investigated topics: Safety, Artificial Intelligence and Autonomous Vehicle. Many synonyms are present in the literature for the terms "artificial intelligence" and "autonomous vehicle". Therefore, an exploratory study of their most representative synonyms was performed. Then, a careful selection of synonyms was made to ensure the search process would have an appropriate coverage. As a result, the following string with Boolean operators was selected: *("safety" AND ("artificial intelligence" OR "machine intelligence" OR "machine learning") AND ("autonomous vehicle" OR "autonomous car" OR "automated vehicle" OR "automated car" OR "self-driven vehicle" OR "self-driving" OR "driverless"))*. Note that the synonyms for each one of the topics are already presented in the Boolean string previously displayed.

Different instances of the search string were created to adapt it to the distinct database search syntax rules, but the same logical value was kept. In each database, the appropriate options were selected to limit the search process to the Title-Abstract-Keyword (TAK) field set. This is an important measure to reduce the number of non-related or duplicated studies retrieved. However, it was observed that not all databases support a search limited on TAK field set, leading to an inflated number of papers found (e.g. SpringerLink). Table I shows the initial number of papers found per database.

TABLE I
NUMBER OF PAPERS PER DATABASE

| Database | #Entries |
|---|---|
| ACM | 36 |
| Engineering Village | 191 |
| ScienceDirect | 81 |
| Scopus | 182 |
| SpringerLink | 3999 |
| Wiley | 329 |
| WoS | 52 |
| **Total** | **4870** |

## *C. Study Selection Criteria and Papers Review*

The study selection process is shown in Figure 2. Each step indicates the number of papers remaining as a sample after the corresponding step was executed. The first selection criterion applied was to ensure that only the studies with the TAK fields returning positive to the Boolean search expression would be selected. The information (metadata) available for each paper found, in the first step of the selection process, was collected by exporting the results to a spreadsheet. A spreadsheet macro was developed to analyze the TAK fields and to properly select the papers. After this check, only 230 papers remained as a sample. Using the spreadsheet Remove Duplicate tools, the duplicated entries were removed. The 97 remaining papers composed the selected sample.

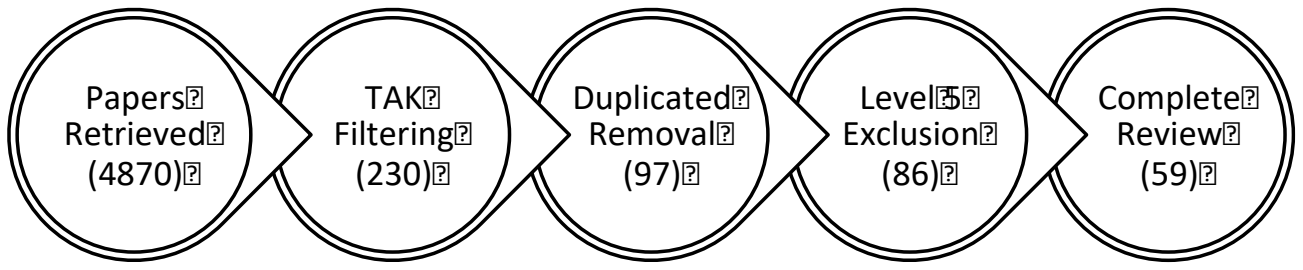

Figure 2. Study selection process.

As a reasonable number of papers (97) was found [10], book chapters, editorials, notes or reports were excluded - level 5 exclusion [10] - and 86 papers remained. The abstracts, titles and keywords of the remaining 86 peer reviewed papers were scrutinized to check their fitness with the goals of this research. After a careful examination (sometimes a full-paper skimming was necessary), 27 papers were considered not related to this research and were excluded from the sample of the literature mapping. Finally, a sample of 59 papers was considered for this study.

There was a considerable drop in the number of studies, from the initial 4870 to the final 59 papers selected. It occurred for different reasons, such as: misuse of the terminology; correct use of the terminology in the context of an example within a paper that did not actually focus on the topic; or lack of restricted search in TAK fields in some databases (in our study, the SpringerLink).

### *D. Data Mapping*

The data mapping from the selected papers were executed after they had been completely reviewed and scrutinized. It was performed categorizing the 59 sample papers into 6 categories (CT.1-6) to answer the 6 research questions (RQ.1-6), respectively. The categories defined were based on the corresponding research question: (CT.1) Impact, (CT.2) Topics, (CT.3) Techniques, (CT.4) Problem, (CT.5) Findings and (CT.6) Future Studies. The categorization process was based on the agreement of researchers working in this study. Different strategies were used to create the codes for each of the categories. For (CT.1) Impact, the code increase was used when the paper described AI as a factor of increasing the safety risk (negative impact on safety) and the code decrease was used when the paper presented AI as a factor of decreasing the safety risk (positive impact on safety). For (CT.2) Topics, (CT.5) Findings and (CT.6) Future Studies, the codifications were derived by the context domain of the study according to what was reported by their authors, as suggested by [11]. Lastly, for (CT.3) Techniques and (CT.4) Problem, similarly to other categories, the codes were based on what was reported by their authors [11] and, due to the wide range of techniques, subfields and misuses of terms, the terminologies were adapted and normalized according to field references [12][13] [14][15][16][17].

## III. Data Analysis

The distribution of the studies over the years can provide an overview of the size and evolution of the field (Figure 3). The left chart in Figure 3 shows the distribution from 1987 until 2018 (April). The oldest study found dates back to 1987. No work was found for over a decade – from 1991 to 2002 – considering the adopted search criteria.

This period can be labeled as the "first winter" in this research topic as an analogy to the Artificial Intelligence "winters"[1]. Only one paper a year was found over the following 3 years – from 2003 to 2005. A second short winter was found from 2006 to 2008. Only 1 paper was found in 2009 and another in 2011, while no paper was found in 2010. Finally, the combination of AI, safety and autonomous vehicles started to get more attention from the scientific community in 2012 when 5 papers were found, although no paper was found in 2013. In fact, 86% (51) of the papers found were published from 2012 to 2018.

The right chart in Figure 3 shows the distribution of the studies over the last decade. The year 2018 was excluded from the plot to avoid misinterpretation. Considering the results presented in Figure 3, the field is gaining momentum based on the continuous growth in the number of published studies since 2014. The trend line built in the last decade data shows a higher angular coefficient, indicating the momentum in recent years.

Most of the papers found are from conference proceedings. In fact, 45 papers (76%) are from conferences. Only 14 papers (24%) were published by journals. Therefore, it is reasonable to expect a growth in the number of publications about this topic in journals. Besides evaluating the time distribution of papers, another important aspect is the consistency-check of the selected keywords in the papers considered. This was performed by checking the most representative keywords among all the synonyms of each of the 3 sets (previously presented) in the search string. All the keywords from the search string found on each paper TAK were accounted. As a result, the total number of hits per keyword was computed. Table II shows the number of studies with each keyword present (hits per keyword) and the percentage of the 59 sample papers with the keyword. Note the sum of the number of hits does not totalize 59. Also, the sum of the percentages for all the keywords for each distinct concept does not totalize 100%. This is because many papers have more than 1 synonym present, which makes it be accounted more than once.

Thus, it is possible to note the most representative keyword for each concept: safety, artificial intelligence and autonomous vehicle. In fact, a search string using only those keywords would result in 36 papers, which corresponds to 61% of the sample size of the present study. However, many other keywords used could not be ignored, since they have a considerable representativeness, such as: machine learning, automated vehicle, self-driving and autonomous car. Conversely the keyword autonomous truck surprisingly had only one hit.

The following sub-sections present the results for each research question (RQ.1-6).

### *A. AI-based systems impact on safety (RQ.1)*

The RQ.1 was answered with the categorization of the sample studies into CT.1 (Impact). Most studies consider AI a technology that increases the system safety (positive impacts on safety). So, 81% (48) of the papers were actually coded as decrease, because they argue that AI decreases the safety risks. Only 19% (11) of the studies consider AI a potential threat to the system safety.

### *B. Main topics of the studies (RQ.2)*

In order to answer RQ.2, the sample papers were classified into the category CT.2 (Topics). Studies were grouped based on their CT.1 coding into two distinct sets: Increase Safety

[1] The Artificial Intelligence field had periods of warm enthusiasms and some periods of very low enthusiasm, with a much lower number of publications and contributions. The literature named those low enthusiasm periods as AI winters.

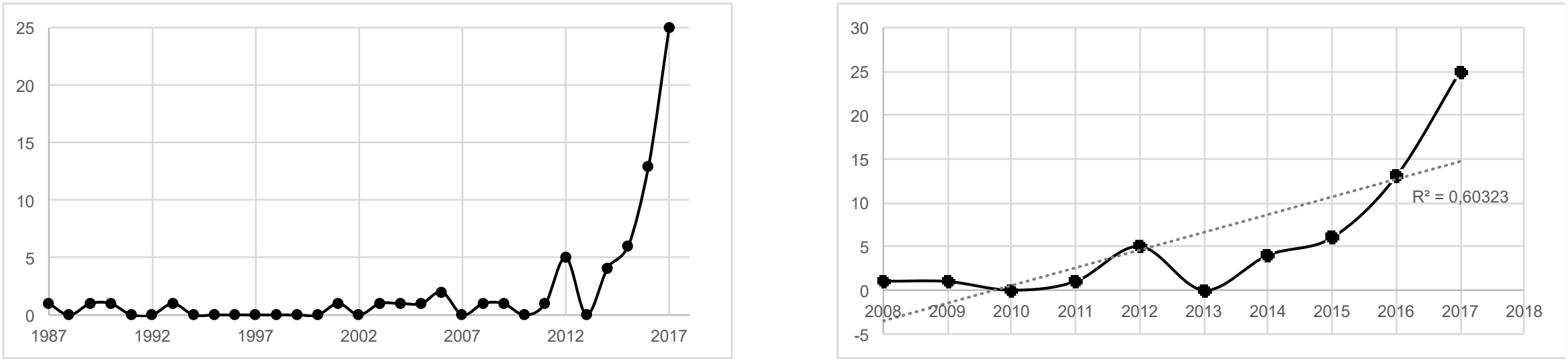

Figure 3. Studies distribution over the years: a) depicting all studies till 2018; b) depicting studies in the last ten years.

TABLE II
KEYWORDS HITS

| Concept | Keyword | #Hits | %Papers |
|---|---|---|---|
| Safety | Safety | 59 | 100 |
| Artificial Intelligence | Artificial intelligence | 36 | 61 |
| | Machine learning | 27 | 46 |
| | Machine intelligence | 1 | 2 |
| Autonomous Vehicle | Autonomous vehicle | 37 | 63 |
| | Automated vehicle | 12 | 20 |
| | Self-driving | 11 | 19 |
| | Autonomous car | 6 | 10 |
| | Driverless | 2 | 3 |
| | Autonomous truck | 1 | 2 |

Risks and Decrease Safety Risks. Then studies were grouped by their similarities and each group was coded with a label that could encompass all its members. Table III shows the results of this coding process. As observed, the papers positioning AI as a factor that decreases safety risks (48 papers, 81%), they studied the subjects related to five main topics: Sensors and Perception (21 papers, 44%), Navigation and Control (13 papers, 27%), Fault Prevention (6 papers, 13%), Conceptual Model and Framework (4 papers, 8%) and Human Factor (4 papers, 8%). In turn, the papers positioning AI as a risk to system safety (11 papers, 19%) studied subjects related to three main topics: Fault Forecasting (5 papers, 45%), Ethics and Policies (4 papers, 36%) and Dependability and Trust (2 papers, 18%). The complete list of references for each code in this category can also be found in Table III.

The main topics for each group of papers differ reasonably from each other. While the papers in the category decrease focus on important aspects to support or to enhance the vehicle autonomy, the papers in the category increase (endanger safety) focus on topics related to safety assurance.

Sensors and Perception is the topic with the largest number of studies (21). They are mostly related to computer vision and detection techniques necessary for adding the necessary capabilities to detect different aspects of the navigation environment and supporting the autonomy of the AVs, such as: general computer vision [19], Doppler sensing [20], lane detection [21], daylight detection and evaluation [22], obstacles detection [23][24][25], pedestrian detection [26][27], pedestrian trajectory prediction [28], road detection [24][29][75], road junction detection [30], road terrain detection [31], traffic signal detection [32][18], situation awareness [33], speed bump detection [25][34], traffic light detection [35], vehicle detection [36] and virtual worlds for training detection [37].

The second largest number of papers (13) found encompasses studies related to Navigation and Control. They are mostly related to techniques necessary to ensure the proper autonomous navigation and control capabilities required by AVs, such as: remote-controlled semi-AV based on IoT [38]; adaptive pre-crash control [39]; safe trajectory selection [76]; AV following another car driven by a human pilot (Trailing) [40]; safe navigation [41]; heuristic optimization algorithm for unsigned intersection crossing [42]; vehicle coordination [43]; maneuver classification [44]; learning to navigate from demonstration [45]; AV movements optimization in intersection [46]; learning and simulation of the Human Level decisions involved in driving a racing car [47]; path tracking [48]; and fuzzy-logic control approach to manage low level vehicle actuators (steering throttle and brake) [49].

Six papers with research related to Fault Prevention were found. These studies encompass researches related to the preventing the occurrence or introduction of faults [50], such as AI for security of wireless communication to ensure safety [51]; remote diagnosis, maintenance and prognosis Framework [52]; prediction of computational workload [53]; vehicle security against cyber-attack [39][54]; and diagnosis of sensor faults [55].

Four studies were found for each of the topics Human Factor and Conceptual Model and Framework. The studies on human factor cover important aspects to be considered in the autonomous cars engineering due to the human-in-the-loop factor, such as: safety, comfort, and stability based on the human driver perception behavior [56]; design of real time transition from assisted driving to automated driving under conditions of high probability of a collision [57]; diagnosing and predicting stress and fatigue of driver in semi-automated vehicles [58]; and advances in driver-vehicle interface [59]. Considering the studies (4) proposing conceptual models and frameworks, they have a considerable diversity of focus, such as: ML and cloud-based framework proposed to address safety and reliability-related issues [60]; AV conceptual model [61]; an interdisciplinary framework to extract knowledge from the large amount of available data during driving to reduce driver's behavioral uncertainties [62]; and a proposition of an AV highway concept to improve highway driving safety [63].

Considering the group of papers positioning AI as a potential factor of decreasing the safety, the highest number of studies was related to Fault Forecasting. In other words, those papers dealt with the limitations to estimate the present number and future incidence of faults in AI-based systems, by executing activities related to evaluation, testing, verification and validation [50], such as: aspects (and limitations) related to safety validation [64]; performance and safety verification methodology [65]; test suites for AV [66]; end-to-end safety for AV design [67]; and a framework to evaluate the impacts of such a sophisticated system on traffic and the impact of continuous increase in the number of highly automated vehicles on future traffic safety and traffic flow [68].

TABLE III
IMPACT OF AI-BASED SYSTEMS ON SAFETY AND ITS MAIN TOPICS AND REFERENCES

| Category | Codes | #Hits | %Papers | References |
|---|---|---|---|---|
| CT.1 - Impact | Decrease Safety Risks (Positive Impact on Safety) | 48 | 81 | |
| CT.2 - Topics | Sensors and Perception | 21 | 44 | [18][75][35][36][21][25][27][29][23][22][34][20][44][32][31][33][26][28][30][24][19] |
| | Navigation and Control | 13 | 27 | [38][39][76][40][41][47][42][43][48][44][45][46][49] |
| | Fault Prevention | 6 | 13 | [39][51][52][53][54][55] |
| | Conceptual Model and Framework | 4 | 8 | [60][61][62][63] |
| | Human Factor | 4 | 8 | [56][57][58][59] |
| CT.1 - Impact | Increase Safety Risks (Negative Impact on Safety) | 11 | 19 | |
| CT.2 - Topics | Fault Forecasting | 5 | 45 | [64][65][66][67][68] |
| | Ethics and Policies | 4 | 36 | [69][70][71][72] |
| | Dependability and Trust | 2 | 18 | [73][74] |

There were four studies related to discussions about Ethics and Policies. One of the studies discussed and performed experiments on how distinct ethical frameworks adopted to make decisions about AV crashes can affect the number of lives endangered [69]. The other studies discuss the scope of AI on AV with ethical aspects [70], ethics in AV design [71], and moral values and ethical principles for autonomous machines [72]. As can be seen, those studies are quite recent since the oldest one was published in 2015.

Finally, 2 papers were found related to Dependability and Trust. Dependability is an important concept in critical systems, because it comprises attributes such as safety, security, availability, reliability and maintainability, as well as how (the mechanisms) to keep these systems attributes [50]. According to [50], trust can be defined as accepted dependability. The studies found are thus related to: safety issues [73] and current mechanisms to ensure robust operation in safety-critical situations facing the introduction of non-deterministic software [74].

### *C. Techniques used (RQ.3) and problems they seek to address (RQ.4)*

Aiming to answer RQ.3 and RQ.4, the sample papers were classified into categories CT.3 (techniques) and CT.4 (problems) based on how their authors described the AI technique used in the study. Then, some terminologies used to define the codification for the categories CT.3 and CT.4 were adapted based on the field literature [12][13][14][15][16][17], when necessary.

Most reviewed papers reported the specific AI related techniques used in the research. Some reported the use of more than one technique, whereas others reported only the approach used. Some papers (14 papers, 24%) were related to general aspects of AI or ML techniques, without mentioning specific techniques used or researched [68][71][72][61][64][67][69][74][70][65][63][58][62][19].

All the techniques found in the reviewed papers were mapped considering the problem (CT.4) that they were solving. As a result, Table VIII - placed in the Appendix - lists the techniques found, the number of papers in which they were used, the main problems they were seeking to address, and the references.

As can be seen, there is a considerable number of studies (22%,13) that used techniques related to artificial neural networks. Also, there is a reasonable number of studies reporting the use of SVM (17%, 10). Some studies used Fuzzy Logic (8%, 5), Bayesian Artificial Intelligence (e.g. Bayesian Deep Learning, Naive Bayes Classifier-NBC, etc) (7%, 4), Hidden Markov Based Models (e.g. Continuous Hidden Markov Model-CHMM and Discrete Hidden Markov Model DHMM) (7%, 4), Estimation Filters (e.g. Kalman Filter and Particle Filters) (7%, 4), Nearest-Neighbour-Based Algorithm (e.g. k-Nearest Neighbours - kNN) (7%, 4), Adaptive Boosting (AdaBoost) (5%, 3), Ramer-Douglas-Peucker or Rameri Douglas algorithm (5%, 3), Haar-like feature detector (5%, 3), Histogram of Oriented Gradient (HOG) (5%, 3), Hough Transformation (5%, 3), Optimization Heuristics (5%, 3), Regression-Based Models (5%, 3) and Principal Components Analysis (PCA) (3%, 2).

Analyzing Table VIII, it shows that each of the following techniques were reported, in all the reviewed papers, only once: Canny Edge Detection Algorithm, Case-based reasoning (CBR), Channel Features, Clustering Algorithm k-mean, Complex Decision Trees (CDT), Conditional Random Fields (CRFs), Distributed Random Forest (DRF), Gaussian Mixture Model (GMM), Linear Temporal Logic (LTL), Local Binary Patterns (LBP), Neuroevolution of Augmenting Topologies (NEAT), Novel Image Recognition Technique, Path Planning Algorithms (A* and D*), Satisfiability Modulo Theories (SMT)

Solver, and Viterbi Algorithm. Thus, there is room for new studies using techniques not yet used or under-represented by the set of papers considered.

### D. *Reported findings (RQ.5)*

Question RQ.5 is answered by CT.5 (findings), based on the information about the findings reported on the sample papers. Some papers did not report specific main findings in a straightforward way because the propose frameworks or approaches had not yet been tested or the results were still incipient. Other papers described very specific findings that would require a background section to support a proper discussion. In those cases, only a higher level of abstraction of the results is presented. Finally, because of space limitation, only some specific examples are described here, while most of the results are presented grouped around the main topic of research. A complete list, oriented by the discussion presented at Section IV, can be found on the Table IX presented in Appendix.

The papers about topics related to Sensors and Perception presented positive and promising results with the techniques employed to address their research problems. In fact, this topic already achieved significant results with the recent developments in AI and sensor technologies. While AI had the image and pattern recognition boosted by advancements such as the new architectures of ANNs and new machine learning techniques, sensor technologies have been boosted in the last decades by the advancements in the robotics and mobile phone industries. As a result, the papers demonstrated applications of enhancements in the techniques or combination of techniques and sensors in order to recognize and to detect important elements and signals the human drivers need to handle to ensure the proper operation of a vehicle. In this context, the findings are positive for the application of ANNs to recognize turn signal [18], road environment and signals [27][32][31][30], and pedestrian [26][28], for example. Likewise, some papers reported SVM has been applied successfully to detect road [75], traffic light [35], and pedestrian [27].

The papers related to Navigation and Control also reported positive and promising results. As presented previously, they used diverse AI techniques to seek to address a broad range of problems. For example, a hybrid AI architecture encompassing ANN, CBR, and a hybrid Case-Based Planner (A* and D* motion planner) was successfully tested to tackle the precrash problem of intelligent control of autonomous vehicles [39], while SVM was used to support a safest path planning in a dynamic environment to avoid maneuvers too close to an obstacle [41].

This SLR found 6 papers for the topic Fault Prevention. Each of these papers used a distinct AI technique for the research problems. One paper presented a preliminary result [53], and another one proposed an approach but did not report results [55]. All the others papers, related to the detection of cyber-attack, presented promising positive results for the application of ANNs [39], Estimation Filters [51], and Fuzzy-Logic [54], for example. Also, preliminary positive results have been reported on the use of a regression-based model to predict the CPU patterns [53].

Two from the four remaining papers related to the topic Human Factor, have presented preliminary positive results. One presented promising results from using a regression-based model to deal with selective attention mechanism [56], while the other presented some examples of scenarios where the use of Bayesian AI could avoid the collision when no action is taken by the human driver [57]. The other 2 papers did not present specific findings, due to their theoretical nature related to the design considerations for the driving assistance system [59] and human drivers monitoring to enhance the integration between AVs and human drivers [58].

The papers proposing conceptual models and frameworks did not present findings related to experimental results. Most of them relied on general AI/ML instead of a specific technique [61][62][63]. Also, besides the proposed approaches themselves, they focused the discussions around the issues they aimed to address, the theoretical background and future potential problems to be addressed in the field.

The last three topics (Fault Forecasting, Ethics and Policies, and Dependability and Trust) have papers more oriented to theoretical discussions and propositions around the challenges AVs are facing or will face related to safety topics, such as test and validation [64], certification [67][74], autonomy assurance and trust when non-deterministic and adaptive algorithms are used [74] - crash assignment facing distinct ethical theories [69], for example. In this context, most of them do not present specific findings using experimental setups; instead, they envision potential future solutions for the discussed challenges. In other words, those papers try to shed an alert light on the important topics that seem to be neglected by the AV enthusiasts, trying to push the research agenda towards safety engineering mindset.

As exceptions, 3 papers presented practical applications and results. [69] presented some interesting findings using a simple experimental simulated environment to test specific crash scenarios under three ethical theories. They found that understanding rational ethics is crucial for developing safe automated vehicles. The results of their experiment indicate that in specific crash scenarios, utilitarian ethics may reduce the total number of fatalities that result from automated vehicle crashes. [66] proposed an approach to describe test-cases for validating autonomous vehicles using recordings of traffic situations for creating a minimal test-suit that could help in the certification process. Considering the example presented, they show how minimalism is achieved by manually comparing the test-cases. Although it is an interesting and promising approach, there are no evidences that it could address a safety certification processes requirement when considering non-deterministic algorithms. Hence, the research was still preliminary. Finally, although [73] presents an end-to-end Bayesian Deep Learning architecture to reduce the risks of hard classifications by adopting probabilistic predictions accounting for each model, no findings from real experiments were presented.

### E. *Reported future studies (RQ.6)*

Question RQ.6 is answered by CT.6 (future studies), based on the collected information about future studies reported on the sample papers. Some papers did not suggest future studies. Other papers described intended future studies or works under development. Those are frequently small incremental changes, such as change of parameter or new test scenarios. Therefore, they are not reported here since their specificities would require a considerable background on the papers contents. That is out of the scope of the systematic literature review.

The studies related to Sensors and Perception propose many future studies, but mostly around improvements that would be made in the future to address some of the limitations of the presented research. Due to the space limitations, only some examples are described here. [18] suggests additional research on image recognition of low contrast images and vehicle images with brake lamps. [35] suggests future work on traffic lights detection under severe weather or night conditions. [34] suggests more research on detecting speed bump during night time. They also suggest research on speed bumper detection when they have no pattern or marking. In addition to that, [34] suggests research to improve the recognition capabilities to distinguish zebra crossing from speed bump. [75] proposed future research about road detection using road lane markers that could be detected by LIDAR, while [21] proposed more research focused on optimizing the lane detection and vehicle recognition algorithms to reduce their computational costs. Also considering the high computational costs, [27] proposed using parallel computing to increase the speed of the image recognition algorithms. Finally, according to [37], additional research is needed on using the virtual environments for testing because the authors believe their usage for training and testing intelligent systems are becoming more relevant.

Most of the studies related to Navigation and Control suggest future studies. The majority suggests extensions to the work they presented. Here, few examples are presented. The study proposing hybrid control architecture [39] suggests an extension to consider the full kinematics and dynamic limitations of the vehicle, while constantly acting to avoid collisions and unsafe driving. The paper proposing an approach using SVM to avoid maneuvers too close to an obstacle by adding a safety margin [41] proposes future re-search to extend it using a combination with the kinetic convex hulls[2] to enable the possibility of computing the solution ahead in time. According to the authors, this would help to predict the position and the width of the optimal margin. As a result, it would improve the approach by adding the ability of reduce the collision risk by preventing the AV from driving into a dangerous situation. The study using Fuzzy Logic as the main approach to control a semi-autonomous car 100-km experiment [40] proposes future research using new sensors and filtering methods for data fusion to reduce the risk on scenarios where the GPS signal is lost. Finally, the study on AVs intersection crossing [46] describes future work in which more types of vehicles and more adjacent intersections would be included in the simulations.

Most of the studies (4 of 6) related to Fault Prevention suggest future studies. Half of the studies are related to security aspects, while the other half is related to diagnosis/prognosis/prediction. The study proposing a cyber-attack detection system based on ANNs [39] suggests a future study to apply the proposed approach to a real vehicle in addition to the application of LSTM to detect online sensor attack. The study proposing the use of Particle Filter and Kalman Filter to secure connected vehicles against DoS attack [51] proposes future work to assess the proposed security scheme under many distinct scenarios, and also to execute tests in real world set-ups. The study about predicting ADAS remaining useful life for the prognosis of its safety critical components using ANNs and other techniques, such as SVM [52], proposes a considerably wide range of future studies, such as using Least Square Support Vector Machine (LSs SVM); using big data techniques to analyze the server data; studying connected vehicle prognosis; using driver, vehicle and region profile data to understand the impact on the environment and driving style impact on the system lifespan; and more studies on prognostics-enabled decision Making (PDM). Finally, the work presenting the use of regression-based methods to predict the CPU usage patterns of software tasks running on an AV [53] suggest future work on the use of some regularization methods for automatic feature selection, but also to particularly investigate the effects of underestimating CPU utilization, and how to handle underestimation of CPU utilization when it happens, aiming to better understand how safe over (or under) estimation of CPU utilization is in terms of reliable autonomous driving.

The studies about Ethics and Policies on AVs basically suggest more research on those topics. In the same way, most of the studies tackling human-factor-related topics do not propose future studies. As an exception, the paper proposing the application of regression-based model for the selective attention mechanism subject [56] proposed a future study to help to reveal the mechanism of rear end collision accident to some extent.

Half (2 of 4) of the studies related to Conceptual Model and Framework do not suggest any future studies. However, implicitly, the next steps would be the deployment of those suggested approaches on experimental set-ups to collect real results. The study proposing a framework to reduce the uncertainty of a driver behavior prediction model [62] suggests more studies focusing on the resilience and sustainability of the system when deployed on a large scale in a complex system.

The papers about Fault Forecasting suggest some future research. Among them, [64] suggests more research on safety envelope mechanisms to describe a boundary within the state space of the AVs rather than trying to prove that it will always work correctly. Koopman, in another paper [67], suggest that the accepted practices must be updated to create an end-to-end design and validation process to address all the safety concerns considering cost, risk, and ethical considerations. [66] proposes more work on creating automated test-cases. [68] proposes more studies based on the framework they proposed to evaluate the impacts of AVs on traffic safety, specially using stochastic simulations with random number seeds to achieve a broader representative and a variety of traffic situations, as well as using the proper statistical analysis techniques to ensure the statistical validity of the results.

[2] Check [76] for more information about kinetic convex hulls.

Finally, the 2 studies about Dependability and Trust also present some suggestions of future studies. [73] asks for more research on new concrete safety evaluation metrics. [74] suggests more research on understanding the dependence of the system components on AVs is needed to establish trust. They also suggest that could be achieved by investigating the many ways in which people, the system, and the environment interrelate.

## IV. SLR Findings Oriented by an AV System Model

In the previous section, the state of the art in the literature about AI on AV safety was identified and investigated by means of a SLR. Six research questions oriented the literature identification, in which studies that include keywords related to safety, AI and AV were considered. The resulting studies were investigated and mapped into 6 categories: Impact (increase or decrease safety risks), Topics (sensors and perception; navigation and control; fault prevention; conceptual model and framework; human factor; etc.), Techniques (general AI/ML; ANN; SVM; etc.), Problem (AV validation; road detection; collision avoidance; etc.), Findings, and Future Studies.

These results considered the AV as a system, but its specific components and functions in an architectural point-of-view were not considered. For deepening the understanding about the state of the art of AI on AV safety it is necessary to show how the presented works are applied/fitted on AV in an architectural point-of-view. In other words, which of AV modules/components and functions are already being developed and which one could be more explored. In order to achieve this goal, it is going to be considered the AV architecture proposed in [78].

An automotive manufacturer consortium (CAMP-AVR) [78] proposed a high-level architecture considering the main system components demanded for the vehicle movement control, to be used in the deployment of future Dynamic Driving Tasks (DDT). Figure 4 (left) illustrates the model considering a traditional vehicle (i.e. human operation with no automation deployed), and Figure 4 (right) illustrates the introduction of some level of machine automation (hybrid) in Sensors, Controller and Actuators elements. While the diagram considering the human operation can solely be mapped to the SAE Automation Level 0 (no automation), the hybrid one encompassing machine automation with human-in-the-loop can be mapped to the SAE Automation Levels 1 to 4 (semi-autonomous) [78].

In this context, a modified version of the semi-autonomous model is proposed here (Figure 5) including the system boundary. Also, the human related components were grouped as one single component (human-in-the-loop), which interacts with Machine Actuators, Machine Control, Machine Perception and Environment. A single component represents a more realistic approach facing the complexity added by the human in the system and allows the examination of the user actions and interactions as suggested by [79]. Also, it supports a necessary human-centered and holistic view [80] to better support the complexity of the human behavior and its interaction to the system. It avoids the misconceptions of the too logical designs from some engineering designs and helps to consider and accept human behavior the way it is, not the way engineers would wish it to be [81]. In fact, this is a necessary upgrade considering the original model is derived by the classical view from the automation engineering for industrial applications, where the environment was under control of the system designer, the human interactions had a considerable narrower scope, and its potential impact to the whole system were much lower, when compared to its application to the semi-autonomous vehicles. As a result, the proposed DDT version (Figure 5) can be used to map the selected scientific literature. Therefore, it can provide a concise perspective on how the field literature covers those main components and which the uncovered areas are. Also, it can provide a good overview on the predominance of the papers valence (increase or decrease) on safety.

Table IV shows how CT.1 (Impact) and CT.2 (Topic) codes are mapped to the components of the modified semi-autonomous system model, as well as the relationship between CT.1 and CT.2. Most of the papers are related to machine perception, followed by papers related to a broad system view. Then, the next largest group of papers is related to the machine control component. The remaining papers are related to the human-in-the-loop aspects. An interesting aspect is that only the studies with a broad system aspect were found to have both CT.1 codes (increase and decrease system safety). Basically, the studies focused on distinct components solely understand AI can increase the safety risk. Therefore, there is a lack of studies with a critical mindset that explore the potential negative impacts of AI on the individual components. Finally, no papers were found related to the vehicle, machine actuators or environment.

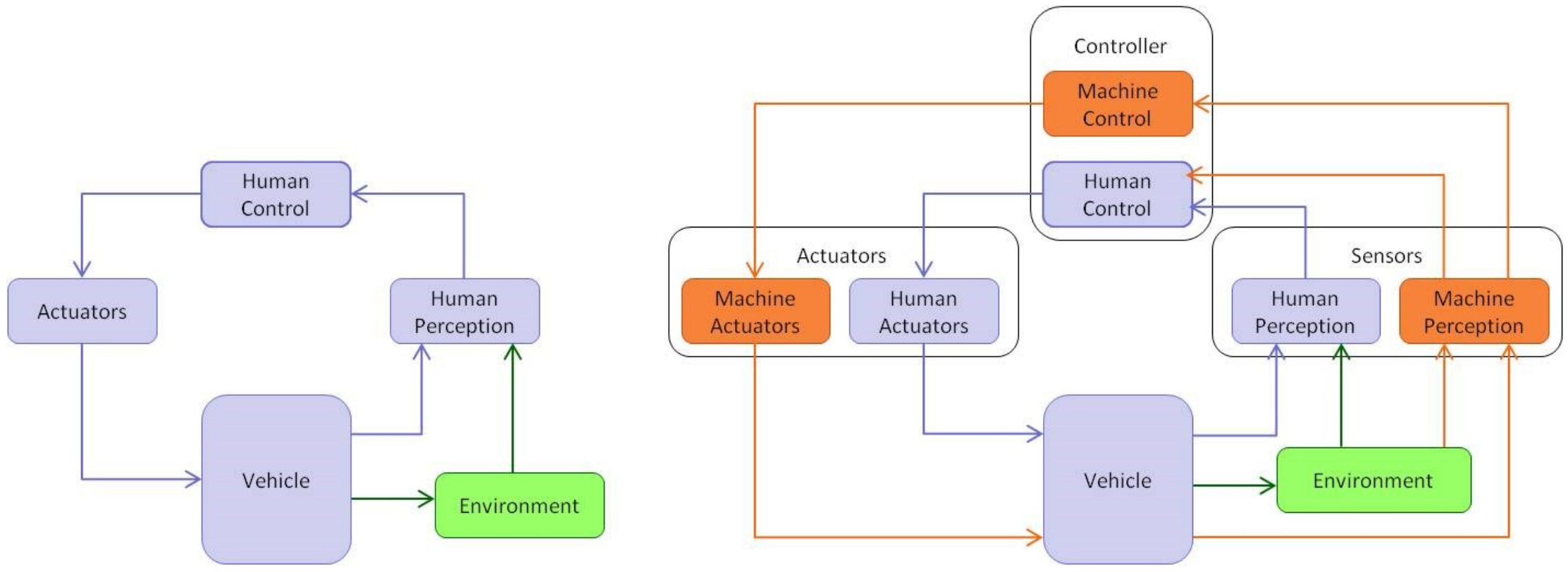

Figure 4. Dynamic Driving Tasks Models: No Automation (left) x Semi-Automation (right) – Source: [78]

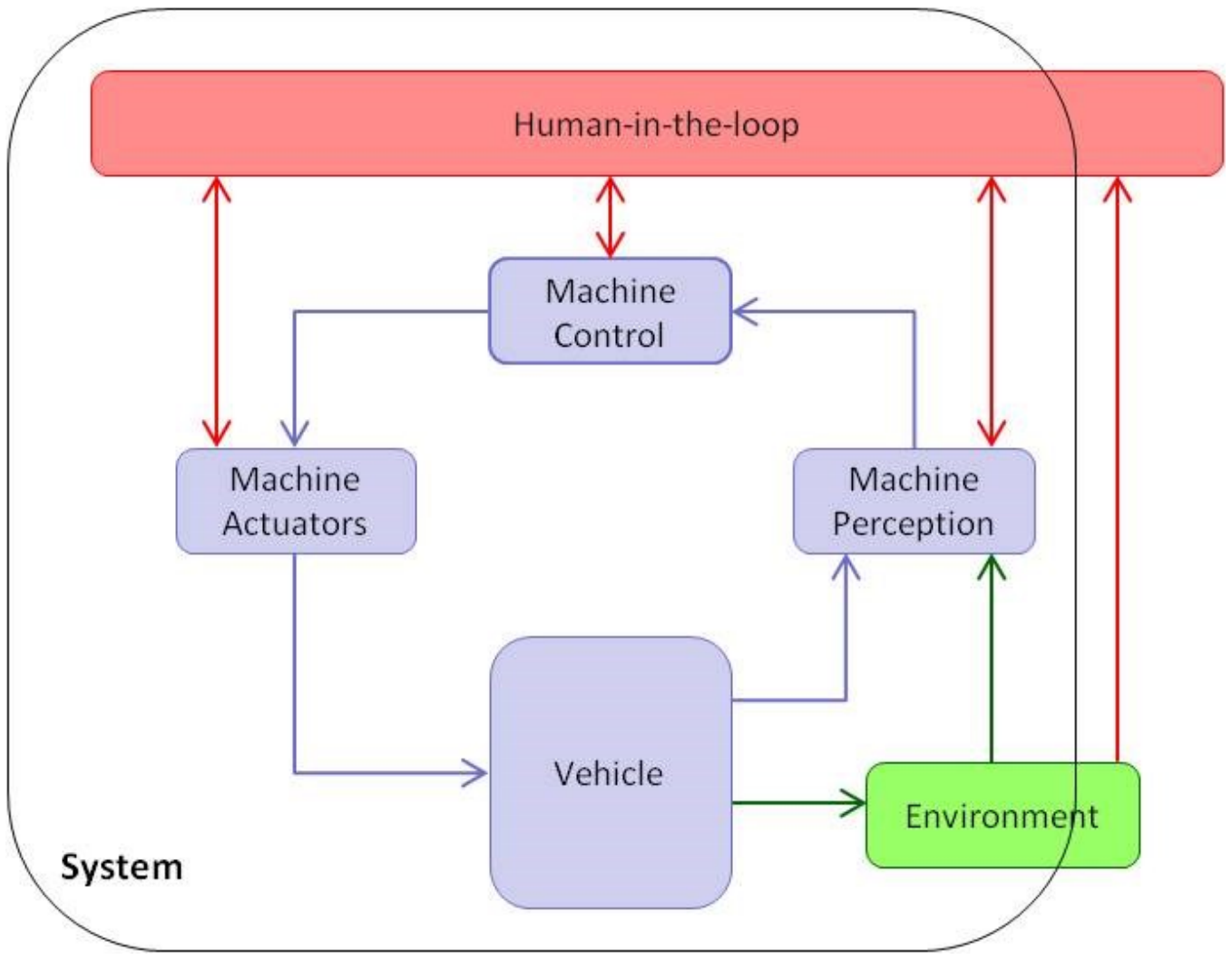

Figure 5. An adapted version of Dynamic Driving Task (DDT) Model

Table V shows how the wide range of AI techniques (CT.3 code) is mapped to the components of the modified semi-autonomous system model. The AI techniques are grouped around their scope: system-oriented (32%, 19) and component-oriented (68%, 40). When a paper uses a combination of techniques, for example, ANN and SVM, it results into a unit added to the total number of papers using ANN and a unit added to the total number of papers using SVM. In this context, most of the studies (63%, 12) related to system-wide scope referred to general AI/MI. Most of the studies (20%, 11) related to machine perception used ANNs. In fact, ANN, SVM and HMM (Hidden Markov Model) account for 48% of the studies related to machine perception. Fuzzy logic (18%, 3) is the most widely used technique in the machine control-related papers. Fuzzy Logic, SVM, Optimization Heuristics and Ramer-Douglas-Peucker or Ramert Douglas algorithm account for 53% of the studies related to machine control. Finally, Bayesian Artificial Intelligence techniques are used in most of the studies (29%) related to human-in-the-loop.

Table VI shows the total count of each AI technique occurrence over the sample papers. The sample papers have different heterogeneity in the applied AI approaches. Besides 24% of the papers using generic AI/ML concepts, 49% of the papers applied only one type of AI technique. Therefore, they are homogeneous in terms of the applied AI technique. In those studies, the most widely used techniques were Artificial Neural Networks (28%, 8), Fuzzy Logic (14%, 4) and Support Vector Machine (SVM) (10%, 3). The remaining 27% employed a hybrid approach by combining multiple types of AI techniques. Among those papers, the combination of Artificial Neural Networks to other techniques (44%, 7), Support Vector Machine to other techniques (SVM) (25%, 4) and Hidden Markov-Based Models (e.g. Continuous Hidden Markov Model-CHMM and Discrete Hidden Markov Model-DHMM) to other techniques (13%, 2) were the most frequent hybrid approaches found in the papers selected.

TABLE IV
MODIFIED SEMI-AUTONOMOUS DDT SYSTEM MODEL X CT3 CODES

| CT.3 - Topic | Component of the Modified DDT System Model (#Hits) | % |
|---|---|---|
| CT.2 – Impact on Safety: Increase Safety (+) | | |
| Sensors and Perception | Machine Perception (21) | 36 |
| Navigation and Control | Machine Control (13) | 22 |
| Human Factor | Human-in-the-loop (6) | 10 |
| Fault Prevention<br>Conceptual Model and Framework | System (+) (8) | |
| CT.2 – Impact on Safety: Decrease Safety (-) | | 32 |
| Fault Forecasting<br>Ethics and Policies<br>Dependence and Trust | System (-) (11) | |

Many different combinations of ANNs with other techniques were found (7 papers). As shown in Table VII most of those papers are related to Sensors and Perception (3 papers) as well as Navigation and Control (2 papers). Also, papers related to Conceptual Model and Framework and Fault Prevention employed hybrid approach (2 papers). The papers that used a combination of models associated to Hidden Markov Based Models were related to Navigation and Control as well as Sensors and Perception. The paper that used Hough Transformation combined to other models is related to Navigation and Control. The paper that employed a combination of techniques to propose a Novel Image Recognition Technique is related to Sensors and Perception. The paper using Regression- Based Models combined to other techniques is related to AV Navigation and Control. Finally, all the papers employing SVM combined to other techniques were related to the topic Sensors and Perception. The same grouping strategy applied to Table V (system-oriented and component-oriented) can be applied here to evaluate the problems (CT.4). ***System-level*** problems include 16 papers: AV Validation [64], Machine-learning-based systems validation to the ultra-dependable levels required for AV [67], Human and Machine Driver Co-existence [60], Coexistence Human Machine Controller [70], Driving Car Tasks Classification [61], Lack of efficient Safety Performance Verification technique when AI/ML is used [65], Crash assignment, especially between automated vehicles and non-automated vehicles [69], Reduce the uncertainty of a driver behavior prediction model [62], Investigate three underexplored themes for AV research: safety, interpretability, and compliance [73], How vehicle autonomy technology can be used to benefit car drivers and also to propose a concept of an autonomous highway vehicle which improves highway driving safety [63], AV decisions in complex dilemmas as a social agent [71], Hybrid (humans and machines) collective decision-making systems [72], Autonomy assurance and trust in Automated Transportation Systems [74], AV Test [66] and, Evaluate the impacts of the number of highly automated vehicles on future traffic safety and traffic flow [68].

TABLE V
TECHNIQUES X DDT SYSTEM MODEL COMPONENT

| DDT System Model | | Technique | #Hits | %Paper | Accum.% |
|---|---|---|---|---|---|
| System-oriented (System) | | General AI/ML | 12 | 63% | 63% |
| | | Hough Transformation related approaches | 2 | 11% | 74% |
| | | Artificial Neural Networks | 2 | 11% | 84% |
| | | Optimization Heuristics | 1 | 5% | 89% |
| | | Estimation Filters (e.g. Kalman Filter and Particle Filters) | 1 | 5% | 95% |
| | | Linear Temporal Logic (LTL) | 1 | 5% | 100% |
| Component-oriented | Machine Perception | Artificial Neural Networks | 11 | 20% | 20% |
| | | Support Vector Machine (SVM) | 8 | 15% | 35% |
| | | Hidden Markov Based Models (e.g. Continuous Hidden Markov Model-CHMM and Discrete Hidden Markov Model-DHMM) | 4 | 7% | 43% |
| | | Estimation Filters (e.g. Kalman Filter and Particle Filters) | 3 | 6% | 48% |
| | | Histogram of Oriented Gradient (HOG) | 3 | 6% | 54% |
| | | Nearest-Neighbor Based Algorithm (e.g. k-Nearest Neighbours - kNN) | 3 | 6% | 59% |
| | | Adaptive Boosting (AdaBoost) | 3 | 6% | 65% |
| | | Principal Components Analysis (PCA) | 2 | 4% | 69% |
| | | Haar-like feature detector | 2 | 4% | 72% |
| | | Fuzzy Logic | 2 | 4% | 76% |
| | | Viterbi algorithm | 1 | 2% | 78% |
| | | Bayesian Artificial Intelligence (e.g. Bayesian Deep Learning, Naive Bayes Classifier-NBC, etc) | 1 | 2% | 80% |
| | | Regression Based Models | 1 | 2% | 81% |
| | | Hough Transformation related approaches | 1 | 2% | 83% |
| | | Ramer-Douglas-Peucker or Ramer-Douglas algorithm | 1 | 2% | 85% |
| | | Novel Image Recognition Technique | 1 | 2% | 87% |
| | | Gaussian Mixture Model (GMM) | 1 | 2% | 89% |
| | | General AI/ML | 1 | 2% | 91% |
| | | Complex Decision Trees (CDT) | 1 | 2% | 93% |
| | | Channel Features | 1 | 2% | 94% |
| | | Local Binary Patterns (LBP) | 1 | 2% | 96% |
| | | Clustering algorithm k-mean | 1 | 2% | 98% |
| | | Conditional Random Fields (CRFs) | 1 | 2% | 100% |
| | Machine Control | Fuzzy Logic | 3 | 18% | 18% |
| | | Support Vector Machine (SVM) | 2 | 12% | 29% |
| | | Optimization Heuristics | 2 | 12% | 41% |
| | | Ramer-Douglas-Peucker or Ramer-Douglas algorithm | 2 | 12% | 53% |
| | | Case-based reasoning (CBR) | 1 | 6% | 59% |

| | | | | |
|---|---|---|---|---|
| | Nearest-Neighbor Based Algorithm (e.g. k-Nearest Neighbours - kNN) | 1 | 6% | 65% |
| | Basic AI Path Planning algorithms such as A* and D* | 1 | 6% | 71% |
| | Artificial Neural Networks | 1 | 6% | 76% |
| | Regression Based Models | 1 | 6% | 82% |
| | Distributed Random Forest (DRF) | 1 | 6% | 88% |
| | Neuroevolution of Augmenting Topologies (NEAT) - ANN + GA | 1 | 6% | 94% |
| | Satisfiability Modulo Theories (SMT) Solver | 1 | 6% | 100% |
| Human-in-the-loop | Bayesian Artificial Intelligence (e.g. Bayesian Deep Learning, Naive Bayes Classifier-NBC, etc) | 2 | 29% | 29% |
| | Regression Based Models | 1 | 14% | 43% |
| | Haar-like feature detector | 1 | 14% | 57% |
| | Canny Edge Detection Algorithm | 1 | 14% | 71% |
| | Hough Transformation related approaches | 1 | 14% | 86% |
| | General AI/ML | 1 | 14% | 100% |

Considering the ***component-level*** problems, 21 papers (36%) are related to dealing with algorithms and techniques to deal with Machine Perception issues, such as: Vehicle Cyber Attack [39], Turn Signal Recognition [18], Securing connected vehicles against Denial of Service (DoS) attack [51], Road Detection [75], Traffic Light Detection [35], Prediction of advanced driver assistance systems (ADAS) remaining useful life (RUL) for the prognosis of ADAS safety critical components [52], Vehicle Detection and Counting [36], predicts the CPU usage patterns of software tasks running on a self-driving car [53], a safety warning and driver-assistance system and an automatic pilot for rural and urban traffic environments [21], reliable and robust obstacles detection continues to be largely investigated and still remains an open challenge, especially for difficult scenarios and, in general cases, with loosened constraints and multiple simultaneous use-cases [25], Pedestrian Detection [27], Road environmental recognition and various object detection in real driving conditions [29], Obstacle clustering and tracking [23]. For an autonomous behavior, each truck must be able to follow the vehicle ahead. Due to that, each vehicle must be able to recognize the leading vehicle [22], Speed bump detection [34], providing road safety to connected drivers and connected autonomous vehicles [20], how to ”automate” manual annotation for images to train visual perception for AVs [44], Road Sign Classification in Real-time [32], Road Terrain detection [31], Spatio-temporal situation awareness [33], Pedestrian detection and movement direction recognition [26], Pedestrian Trajectory Prediction [28], Road junction detection [30], Cyber Attack in V2X [54], Learn from Demonstration [45], Early detection of faults or malfunction [55], Road and Obstacle Detection [24] and Enhance Image Understanding [19].

The problems related to Machine Control were found in 17 papers (22%). Those problems include: Pre-Crash problem of Intelligent Control of autonomous vehicles robot [39], Safe-optimal trajectory selection for autonomous vehicle [76], Driverless car 100 km experiment [40], Robot maneuvers too close to an obstacle, which increases the probability of an accident. Preventing this is crucial in dynamic environments, where the obstacles, such as other UAVs, are moving [41], Learning and simulation of the Human-Level decisions involved in driving a racing car [47], Control intersection crossing (all way stop) and optimizing it [42], How to prove the correctness of an algorithm for Vehicle Coordination [43], Path tracking [48], Drivers maneuver classification [44], AVs intersections crossing optimization [46] and Manage low level vehicle actuators (steering throttle and brake) [49].

Finally, the problems related to Human-in-the-loop new DDT component (Figure 5) are present in 7 papers (10%). Those problems include: Selective Attention Mechanism [56], Developing remote controlled car with some automation to deal with traffic light detection, obstacle avoidance system and lane detection system to be driven from anywhere over a secured internet connection [38], Collision avoidance when no action is taken by driver to avoid the collision [57], Human drivers monitoring system to ensure they will be able to take over control within short notice [58] and, Design of driving assistance system [59]. This seems to be an attention-point; this problem category can be considered one serious challenge to semi-autonomous vehicles (SAE Level 1 to Level4). Therefore, more research is needed into this topic because only 6 papers were found.

TABLE VI
HETEROGENEITY OF THE USED AI APPROACHES

| Heterogeneity | % | Main Technique | #Hits | %Papers |
|---|---|---|---|---|
| Generic | 24% | General AI/ML | 14 | 100% |
| Homogenous | 49% | Artificial Neural Networks | 8 | 28% |
| | | Fuzzy Logic | 4 | 14% |
| | | Support Vector Machine (SVM) | 3 | 10% |
| | | Regression Based Models | 2 | 7% |
| | | Estimation Filters (e.g. Kalman Filter and Particle Filters) | 2 | 7% |
| | | Bayesian Artificial Intelligence | 2 | 7% |
| | | Optimization Heuristics | 2 | 7% |
| | | Ramer-Douglas-Peucker or Ramer-Douglas algorithm | 2 | 7% |
| | | Hough Transformation | 1 | 3% |
| | | Satisfiability Modulo Theories (SMT) Solver | 1 | 3% |
| | | Adaptive Boosting (AdaBoost) | 1 | 3% |
| | | Linear Temporal Logic (LTL) | 1 | 3% |
| Hybrid | 27% | Artificial Neural Network combined to other techniques | 7 | 44% |
| | | Support Vector Machine (SVM) combined to other techniques | 4 | 25% |
| | | Hidden Markov Based Models (e.g. Continuous Hidden Markov Model-CHMM and Discrete Hidden Markov Model-DHMM) combined to other techniques | 2 | 13% |
| | | Hough Transformation related approaches combined to other techniques | 1 | 6% |
| | | Regression Based Models combined to other techniques | 1 | 6% |
| | | Novel Image Recognition Technique | 1 | 6% |

## V. Final Remarks

Machine Perception has more studies with practical results. Considering the other components, few studies with practical results from real deployments were found. Most of the papers presented preliminary results. In fact, some papers start with a promise and finish with more promises. Considering only 24% of the total papers considered in this study were published by journals, it is possible to conclude the field is not mature yet.

Some similar issues were studied in more than one paper about Machine Perception, and distinct techniques were applied to address them (for example, ANN and SVM applied to similar issues as well issues as well distinct techniques applied to the topic cyber-attack). Considering some of those techniques have different working mechanisms, that fact can be an important finding for the safety of autonomous cars as regards the need of redundant components. Similar issues being addressed by different techniques were not identified for Machine Control.

The papers related to Navigation and Control also reported positive and promising results, although the level of maturity of the achievements are clearly much lower than the sensors and perception as well as far from what would be expected for an autonomous vehicle considering the potential hazardous situations it may face. In fact, most of the results presented are preliminary.

Only few of the studies related to system described practical results from real deployments. The papers proposing conceptual models and frameworks bring important contributions, but they are mostly not tested in real set-ups. There is thus a lack of reported results from models and frameworks that could build the foundation of AVs safety. Also, few human-in-the-loop studies had practical results from real deployments. However, they seem to be one of the most important topics seeing that there will be more semi-autonomous cars than fully autonomous ones for a while, and they will co-exist. The human factor will thus be an important variable in the system to be considered not only as the impacted side of the safety, but as one of the sources of interactions influencing the safety levels. The topic requires multidisciplinary studies involving fields beyond engineering and computer science, such as neurosciences. This shows the field is not mature yet.

The amplitude and range of the reported future researches in the papers reviewed suggest that there is an empty space for new research into this field. For example, only few studies were found about the three topics positioning AI as a potential source of negative impact on safety - Fault Forecasting, Ethics and Policies, and Dependability and Trust. When combined to the other findings reported by the present study, it confirms the impressions formed during an exploratory research of the literature [1]. It reinforces the perception that the field of AI and AV is not heavily influenced by the safety engineering culture yet. In fact, the studies published about this current topic seem to be more driven by computation-related domains, with no tradition regarding safety culture, than other fields that are much more connected to safety in critical systems [1].

Additional research is necessary for most of the studies reviewed. They need to be extended to be tested in simulated or real set-ups, new and broader scenarios, with new and more data, and consider experimental designs whereby the results from the proposed approach are compared to benchmarks and alternative techniques. Many AI techniques have achieved impressive results; however, it is still arguable whether the error rates are suitable for real deployments in AVs under the light of a (missing) hazard analysis. Therefore, additional studies with improvements in those techniques are required. Finally, a stronger influence of safety engineering on most of the studies would benefit the field.

A research agenda must consider a serious safety agenda for future studies, at system-level, component-level and AI technique-level. In fact, there are some topics related to safety concerns over AVs, which are critical-path to the development of the field. Some of the suggested topics are related to the challenges with validating machine-learning- based systems to the ultra-dependable levels required for AVs; wider and deeper studies about human-machine collaboration in the context of AVs; autonomy assurance and trust in AVs; ethical and moral decisions in the context of AVs; among other topics, from Validating machine-learning-based systems to the ultra-dependable levels required for AVs; and autonomy assurance and trust in AVs seem to be the holy grail towards a fully autonomous AV - SAE level 5 . They are also key topics for the Safety Certification of non-deterministic control systems. In those contexts, there are many gaps to be filled by future researches, such as AVs software testing, Fault Injection Testing for AI on AVs, Failure Modes and Effects Analysis (FMEA) for AI on AVs, AI safeguards for AVs, AI safety envelopes for AVs, AI redundancy for AVs (many possible approaches, such as a hybrid connectionist and symbolic architecture using causal inference), explainable AI for AVs, AI fault forecasting. Finally, studies on V2X communication can help autonomy assurance by providing channels for hardware and software redundancy. Human-machine collaboration in the context of AVs is another key topic with special impact on the semi-autonomous vehicles (SAE levels 1 to 4). Investigations on the best way humans and AVs can interact during normal operations and facing hazardous situations are needed to meet the adequate safety requirements the semi-autonomous vehicles must have. Those studies must consider hybrid collective decision-making systems to enable humans and machines to work together and to agree on common decisions, as well as how to deal with the lack of agreement in some situations.

Also, there is another important discussion arising in the context of human-machine collaboration that must be investigated. On the one hand, there are reports about advanced driver assistant technologies that failed (such as Tesla Autopilot) and the driver was not able to react in time to avoid the accident. They ended-up in life losses and property damages. On the other hand, there are reports about situations in which the advanced driver assistant technologies saved the drivers' life by automatically taking the driver suffering a heart attack to the hospital; fully controlling the car with a drunk driver sleeping; and using a defensive lane change maneuver to avoid being hit by a truck changing its lane. Some players in the industry are pushing the automation evolution steps towards full automation by requiring the human driver to be a backup to the automated driver. Other players in the industry believe the automated driver must be a backup to the human driver. It looks like the second approach can be a smoother and safer path towards a SAE level-5 automation.

TABLE VII
HYBRID AI APPROACHES X TOPIC

| Main AI Technique | Topic | AI Techniques | Reference |
|---|---|---|---|
| Artificial Neural Network | Conceptual Model and Framework | HoughTransforms, HoughLines, LocalMaximaFinder, Kalman filters and Convolutional Neural Network (CNN) | [60] |
| | Fault Prevention | KNN, SVM Regression (SMO), ANN | [52] |
| | Navigation and Control | CBR, ANN, fuzzy logic, Nearest-Neighbor Retrieval Algorithm, Basic AI Path Planning algorithms such as A* and D* | [39] |
| | | ANN combined to Genetic Algorithm - Neuroevolution of Augmenting Topologies (NEAT) | [47] |
| | Sensors and Perception | ANNs, AdaBoost, SVM, Hidden Markov Models (HMMs), CRFs | [30] |
| | | Clustering algorithm k- mean, ANN | [31] |
| | | HOG, SVM, PCA, ANN | [29] |
| Hidden Markov Based Models | Navigation and Control | GMM, Continuous Hidden Markov Model (CHMM), Discrete Hidden Markov Model (DHMM) | [45] |
| | Sensors and Perception | HMM, Viterbi algorithm, Adaboost trained Haar-like feature detector | [36] |
| Hough Transformation | Navigation and Control | Haar Feature Based Cascade Classifier, Canny edge detection and Hough line transformation | [38] |
| Novel Image Recognition Technique | Sensors and Perception | Combination of mathematical techniques | [34] |
| Regression Based Models | Navigation and Control | (DRF) and Linear Regression (LR) | [76] |
| Support Vector Machine (SVM) | Sensors and Perception | Haar, HOG, LBP, Chanel features, SVM | [37] |
| | | k-Nearest Neighbours (kNN), Naïve Bayes classifier (NBC), SVM | [27] |
| | | Principal component analysis network (PCANet), SVM | [35] |
| | | SVM, HOG | [75] |

Immersive environments for training and testing AVs represent another research trend. As the underlying technologies supporting AVs development evolve, higher automation-levels become possible. Considering the potential hazards until the AVs are well trained and fine-tuned, the immersive technologies are becoming an important tool to support the development, training and tests of fully autonomous machines. Another broad topic requiring further research is related to ethical and moral decisions in AVs. Some studies only mention issues related to moral dilemmas while others provide some simple experiments involving simulated environments and/or human interviews. However, they misinterpret important concepts and bring the discussions around the decisions AVs must make when life losses are involved, besides the moral and ethical perceptions from the human perspective. All of them miss important points such as statistical considerations and the societal result. In other words, the discussions are not deep enough as regards situations such as whether an AV should hit an old man or a child, while a true safe machine control should consider all the probabilities involved and select the one that minimizes the chances of life losses instead of just picking an option. For example, the system must consider small signals, such as which of the potential victims is paying attention to the approaching AV and what would their potential reaction be and effectiveness of it based on the age and other metrics, as well, considering the multiple scenarios, and the configuration of each, such as speed, region of the car hitting which region of each victim, the potential damages and the severity of the damages considering the estimated weight and overall physical condition, to decide based on the minimization of chances of life losses. This approach will result into higher safety levels for society. Finally, only 1 paper about autonomous truck was found. Considering some specificities of autonomous truck and its risks, at least a few more studies about the topic could be expected.

## APPENDIX

TABLE VIII
CT.3 X CT.4

| Technique (CT.3) | Hits | Papers | Addressed Problem (CT.4) | References |
|---|---|---|---|---|
| General AI/ML | 14 | 24% | AV Validation; Challenge with validating machine-learning based systems to the ultra-dependable levels required for autonomous vehicle; Coexistence Human Machine Controller; Driving Car Tasks Classification; Lack of efficient Safety Performance Verification technique when AI/ML is used; Crash assignment, especially between automated vehicles and non-automated vehicles; Reducing the uncertainty of a driver behavior prediction model; Integration between automatic vehicle and human driver; How the vehicle autonomy technology can be used to benefit car drivers and to improve highway driving safety by a concept of an autonomous highway vehicle; AV decisions in complex dilemmas as a social agent; Hybrid (humans and machines) collective decision making systems (work together and agree on common decisions); Autonomy assurance and trust (CERTIFICATION PROA CESS) in Automated Transportation Systems; Evaluating the impacts of the number of highly automated vehicles on future traffic safety and traffic flow; Enhancing Image Understanding. | [64],[67], [70],[61], [65],[69], [62],[58], [63],[71], [72],[74], [68],[19] |
| Artificial Neural Networks (ANN) | 13 | 22% | Vehicle Cyber Attack; Turn Signal Recognition; Pre-crash issues of Intelligent Control of autonomous vehicles robot; Real-time Road Sign Classification; Road Terrain detection; Spatio-temporal situation awareness; Pedestrian detection and movement direction recognition; Pedestrian Trajectory Prediction; Road junction detection; Early faults or malfunction detection; Prediction of advanced driver assistance systems (ADAS) remaining useful life (RUL) for the prognosis of ADAS safety critical components; Road environmental recognition and various objects detection in real driving conditions; Human and Machine Driver Co-existence; | [39],[18], [32],[31], [33],[26], [28],[30], [55],[52], [29],[60] |
| Support Vector Machine (SVM) | 10 | 17% | Road Detection; Robot maneuvers too close to an obstaR cle; Road environmental recognition and various object detection in real driving conditions; Drivers maneuver classification; Traffic Light Detection; | [75],[41], [29],[20], [44],[35], |
| | | | Prediction of adc vanced driver assistance systems (ADAS) remaining useful life (RUL) for the prognosis of ADAS safety critical components Pedestrian Detection; How to ”automate” manual annotation for images to train visual perception for AVs Road junction detection; | [52],[27], [37],[30] |
| Bayesian Artificial Intelligence | 4 | 7% | Collision avoidance when no action is taken by driver; Safety, interpretability, and compliance; Pedestrian Detection; Design of driving assistance system; | [57],[73], [27],[59] |
| Fuzzy Logic | 5 | 8% | PreCrash problem of Intelligent Control of autonomous vehicles robot; Driverless car 100 km experiment Cyber Attack in V2X; Manage low level vehicle actuators (steering throttle and brake); Road and Obstacle Detection; | [39],[40], [54],[49], [23] |
| Hidden Markov Based Models | 4 | 7% | Vehicle Detection and Counting; Road junction detection; Learn from Demonstration; | [36],[29], [45] |
| Estimation Filters | 4 | 7% | Human and Machine Driver Co-existence; Securing connected vehicles against Denial of Service (DoS) attack; Reliable and robust obstacles detection; | [60],[51], [24] |
| Nearest Neighbour-Based Algorithm | 4 | 7% | Pre-crash problem of Intelligent Control of autonomous vehicles robot; Pedestrian Detection; Providing road safety to connected drivers and connected autonomous vehicles; | [39],[26], [19] |
| Adaptive Boosting (AdaBoost) | 3 | 5% | Vehicle Detection and Counting; Leading vehicle recogV nition in platooning; Road junction detection; | [36],[21], [29] |
| Ramer-Douglas Peucker or Ramer-Douglas algorithm | 3 | 5% | Obstacle clustering and tracking; Path tracking; | [22],[48] |
| Haar-like feature detector | 3 | 5% | Developing remotecontrolled car with some automation to deal with traffic light detection, obstacle avoidance system and lane detection system to be driven from anywhere over a secured internet connection; Vehicle Detection and Counting; How to "automate" manual annotation for images to train visual perception for AVs; | [38],[36], [37] |
| Histogram of Oriented Gradient (HOG) | 3 | 5% | Road Detection; Road environmental recognition and various objects detection in real driving conditions; How to "automate" manual annotation for images to train visual perception for AVs; | [75],[28], [37] |
| Hough Transfor-mation | 3 | 5% | Road Detection; Road environmental recognition and various object detection in real driving conditions; How to "automate" manual annotation for images to train visual perception for AVs; | [60],[38], [20] |
| Optimiza-tion Heuristics | 3 | 5% | Control intersection crossing (all way stop) and op- timization; Autonomous vehicles intersections crossing optimization; Human and Machine Driver Co-existence; | [46],[42], [60] |
| Regression-Based Models | 3 | 5% | Selective Attention Mechanism; Safe-optimal trajectory selection for autonomous vehicles; Predicts the CPU usage patterns of software tasks running on a self-driving car; | [56],[79], [53] |
| Principal Componen-ts Analysis (PCA) | 2 | 3% | Traffic Light Detection; Road environmental recognition and various object detection in real driving conditions; | [35],[28] |
| Canny Edge De- tection Algorithm | 1 | 2% | Developing remote- controlled car with some automation to deal with traffic light detection, obstacle avoidance system and lane detection system to be driven from anywhere over a secured internet connection; | [38] |
| Case-based reasoning (CBR) | 1 | 2% | Pre-crash problem of Intelligent Control of autonomous vehicles robot; | [39] |
| Channel Features | 1 | 2% | How to ”automate” manual annotation for images to train visual perception for AVs; | [37] |
| Clustering Algo- rithm k-mean | 1 | 2% | Road Terrain detection; | [30] |
| Complex Decision Trees (CDT) | 1 | 2% | Providing road safety to connected drivers and connected autonomous vehicles; | [19] |
| Conditional Random Fields (CRFs) | 1 | 2% | Road junction detection; | [29] |
| Distributed Random Forest (DRF) | 1 | 2% | Safe-optimal trajectory selection for autonomous vehicle; | [79] |
| Gaussian Mixture Model (GMM) | 1 | 2% | Learn from Demonstration; | [45] |
| Linear Temporal Logic (LTL) | 1 | 2% | AV Test; | [66] |
| Local Binary Patterns (LBP) | 1 | 2% | How to "automate" manual annotation for images to train visual perception for AVs; | [37] |
| Neuroevo-lution of Augmen-ting Topologies (NEAT) | 1 | 2% | Learning and simulation of the Human-Level decisions involved in driving a racing car; | [47] |
| Novel Image Recogni-tion Technique | 1 | 2% | Speed bump detection; | [34] |
| Path Planning Al- gorithms (A* and D*) | 1 | 2% | PreCrash problem of Intelligent Control of autonomous vehicles robot; | [39] |
| Satisfiability Modulo Theories (SMT) Solver | 1 | 2% | How to prove the correctness of an algorithm for Vehicle Coordination; | [43] |
| Viterbi Algorithm | 1 | 2% | Vehicle Detection and Counting; | [36] |

TABLE IX
FINDINGS ON PAPERS ORIENTED BY THE DISCUSSION

| **Issue** | Suggested Approach | **AI Technique** | Findings | **Reference** |
|---|---|---|---|---|
| DDT System Model Component - System | | | | |
| Human and Machine Driver Co-existence | Continuously monitor the driving behavior of the neighboring vehicles, sensor behavior and processor behavior of the ego vehicle regardless of the vehicle being autonomous or not | ANN (Hybrid) | The authors have foreseen and proposed the solutions for future problems, which would occur while the autonomous vehicles are a part of driving. All the three described architectures have addressed the safety related problems. The architectures are based on the availability of the resources for vehicles. The first two architectures address the safety failure due to the human ignorance and autonomous vehicle behavior. Third architecture addresses the way of securing the failed vehicle due to system failure. All the architectures rely highly on the efficient connectivity and | [60] |
| | | | computer vision algorithms. | |
| Investigate three under-explored themes for AV research: safety, interpretability, and compliance | End-to-end Bayesian deep learning architecture to propagate uncertainty throughout the AV framework. In this case, standard deep learning makes hard predictions, whereas Bayesian deep learning outputs probabilistic predictions accounting for each model's ignorance about the world. | ANN (Hybrid) | 3 critical themes for a smooth adoption of AV systems by society were highlighted. Hard decisions are dangerous. Soft (uncertain) classifications should be propagated through to the decision layer. This enables the AV to act more cautiously in the event of greater uncertainty. We also discussed the themes. Authors argument interpretability and compliance as ways to build trust and mitigate fears which passengers might otherwise reasonably have about unfamiliar black-box AV systems. Also, they discussed about the importance of clear metrics to evaluate each component's probabilistic output based on their ultimate effect on the vehicle's performance. | [73] |
| AV Validation | Safety Envelopes | General AI/ML | No findings, only suggested approaches. | [64] |
| Challenge with validating machine-learning based systems to the ultra-dependable levels required for autonomous vehicle | Safety certification strategy addressing the cross-disciplinary concerns of safety engineering, hardware reliability, software validation, robotics, security, testing, human-computer interaction, social acceptance, and a viable legal framework. | General AI/ML | The paper only points out the challenge with validating machine-learning based systems to the ultra-dependable levels required for autonomous vehicle fleets, and how that challenge relates to a number of other areas. It does not provide any particular finding. | [67] |
| Coexistence Human Machine Controller | In-car Virtual Assistants | General AI/ML | It is too early to assess whether carmakers' optimistic vision of in-car virtual assistants will match users' experience or follow a similar destiny of "Clippy the Paperclip" from Microsoft Word. Proprietary systems are predominant and there is no common framework addressing ethical principles, liability, data protection, privacy and security on many of the technologies associated to AVs. Due to these shortcomings, allowing a fully autonomous approach could generate more drawbacks than benefits. At least in a first phase, it seems more appropriate to apply AI-based virtual assistants to autonomously execute tasks and take decisions (i.e. replace humans) on safety-related functions only, in line with the requirements defined by international standards. | [70] |
| Driving Car Tasks Classification | Classification of the tasks that take place during the driving of the vehicle and its modeling from the perspective of traditional control engineering and artificial intelligence | General AI/ML | The major issues realted to safety and and the efforts to make sure the technologies involved are robust are discussed: test the safety of the ADAS, standardization, the development of models and algorithms, the appropriate constructing solutions for implementation, and ethical issues. No specific findings are presented. | [61] |
| Lack of efficient Safety Performance | Methodology to generate an estimation of | General AI/ML | Detailed methodology was proposed to deal with the issue by means of statistical | [65] |
| Verification technique when AI/ML is used | probability of fatal mishap of an autonomous UGV navigation algorithm based on Statistical Testing in a MonteCarlo manner in a Simulated Environment | | testing via simulation. Demonstration was still a work in process. | |
| Crash assignment, especially between automated vehicles and nonautomated vehicles | The integration of three ethical theories—utilitarianism, respect for persons, and virtue ethics—with vehicle automation is used in a simple crash scenario where an automated vehicle must choose between three crash types on the basis of a randomly assigned ethical theory to understand the outcomes of distinct ethical frameworks | General AI/ML | The results of the experiment indicated that in specific crash scenarios, utilitarian ethics may reduce the total number of fatalities that result from automated vehicle crashes, although other ethical systems may be useful for developing rules used in machine learning. The experiment demonstrates that understanding rational ethics is crucial for developing safe automated vehicles. | [69] |
| Reduce the uncertainty of a driver behaviour prediction model | Proposes a Data Analysis Framework to exploit AI, quantified self, internet of things and automated driving to build a computational driver behavioural model aming to reduce the uncertainty of a driver behaviour prediction model and be used monitor, predict and control a transportation system. | General AI/ML | It is very hard to predict due to the fluidity and interactions of the driving factors determining the driver performance. | [62] |
| How vehicle autonomy technology can be used to benefit car drivers and also to propose a concept of an autonomous highway vehicle which improves highway driving safety | Conceptual discussion | General AI/ML | No specific findings were presented by the conceptual discussion | [63] |
| AV decisions in complex dilemmas as a social agent | Proposition of a framework for an ethics policy for the artificial intelligence of an AV | General AI/ML | The ethics of automated vehicles is becoming a major issue from legal, social, and vehicle control perspectives. AV will have to make decisions that might eventually harm an agent and that these decisions should not contradict the interests of the end users or the principal stakeholders. An ethics policy for automated vehicles is needed and the proposed framework (AVEthics) is only the beginning of a long path. | [71] |
| Humans and machines will often need to work together and agree on common decisions. | Shared moral values and ethical principles | General AI/ML | Hybrid collective decision-making systems will be in great need | [72] |
| Autonomy assurance and trust (certification process) in Automated Transportation Systems | Framework Proposition for the discussion around the topic | General AI/ML | Authors explored some of the unique challenges that autonomous transportation systems present with regard to traditional certification approaches such being non-deterministic and employing adaptive algorithms. Authors discussed the concept of multiparty trust and how it can be extended to a framework illustrating the relationships between disparate roles. Two thought experiments showed that building and maintaining trust in the perception and judgment of increasingly autonomous systems will be a challenge for the transportation community. | [74] |
| Evaluate the impacts of the number of highly automated vehicles on future traffic safety and traffic flow | Framework to evaluate the impacts of AV on traffic and the impact of continuous increase in the number of highly automated vehicles on future traffic safety and traffic flow | General AI/ML | The results of impact evaluation in this study show that the increase in the penetration rate of the highly automated vehicle together with proper adjustment of model parameter may result in considerate improvements of safety in traffic in terms of defined indicators. The developed methodology allows to compare traffic efficiency and safety measures with different penetration rates in various scenarios by means of microscopic traffic simulation. | [68] |
| AV Test | Creation of Minimal Test-Suites with Test-cases for the validation of AVs using recordings of traffic situations | LTL | The process of test-case derivation can be applied was demonstrated. According to the authors, the derivation of test-cases categorizes the recordings automatically and allows test engineers to specify test inputs. The test-case descriptions use the Linear Temporal Logic (LTL) and allow the execution of continuous behaviors, which may also react to the behavior of the tested vehicle. According to the authors, as the traffic recordings can also be used for machine learning algorithms, the contributes to the discussion of their safety certification. They also state the approach is flexible as it can be extended if new traffic situations are supposed to be covered by testing. | [66] |
| DDT System Model Component – Human-in-the-loop | | | | |
| Collision avoidance when no action is taken by driver to avoid the collision | Real time transition from assisted driving to automated driving under conditions of high probability of a collision if no action is taken to avoid the collision | Bayesian Artificial Intelligence | Systems can be designed to feature collision warnings as well as automated active safety capabilities. The high-level architecture of the Bayesian transition model seems promising. Example scenarios illustrate the function of the real-time transition model. | [57] |
| Design of driving assistance system | Discussion about Design considerations are advanced in order to | Bayesian Artificial Intelligence | No specific findings, only discussions about the important considerations to be taken into account when designing AVs | [59] |
| | overcome issues in in-vehicle telematics systems | | | |
| Integration between AV and human driver | un-obstructive human driver monitoring approaches to ensure they will be able to take over control within short notice | General AI/ML | One of the most essential parts of the autonomous driving system is to monitor driver's physical and mental state to avoid unexpected traffic accidents. There are some solutions for non-contact measurement of vital signs, such as HR, RR include laser Doppler, microwave Doppler radar, and thermal imaging. Several AI approaches that have been applied in classifying non-contact physiological sensor signals in different other domains could be possible to investigate in classifying driver's signals. This paper shows that the assessment of non-contact physiological parameters presents a greater challenge and few attempts have been made to adopt it for the driving situation. | [58] |
| Develop remote controled car with some automation to deal with traffic light detection, obstace avoidance system and lane detection system to be driven from anywhere over a secured internet connection | Traffic Light Detection: Haar Feature Based Cascade Classifier; Lane Detection: Canny Edge detection and Hough line transformation was used | Hough Transformation related approaches combined to other techniques | Low-cost remote-controlled car prototype with basic automated functions and using basic off-the-shelf components with promising results. It proved cheap and useful prototypes can be built for research. Results with the proposed road detection techniques proved to be highly efficient for a road with clearly visible lane market. Canny Edge detection proved to have low error rate. | [38] |
| Selective Attention Mechanism | Weber–Fechner law | Regression Based Models | Besides, the model is consistent with the famous Weber–Fechner law. The Weber–Fechner law says that all people's feeling, including visual feeling, auditory feeling and so on all comply with the fact that the feeling is not proportional to the strength of the corresponding physical quantity but proportional to the logarithm of the corresponding physical quantity. | [56] |
| DDT System Model Component – Machine Control | | | | |
| Pre-crash problem of AV Intelligent Control | AV Intelligent Adaptive Control architecture using an hybrid AI architecture: CBR Engine for Adaptive control (high level) + hybrid Case-Based Planner (A* and D* motion planner) | Artificial Neural Network combined to other techniques | Its is flexible to be integrated to lower levels of vehicles controller and path planners as (A* & D*). It is an ongoing reasearch. Some limited and embrionary experimental are mentioned and the authors claim present research ideas for different pre-crash scenarios and cases such as intersection safety and some general cases. The paper also discusses approaches to integrate basic kinematics of AVs features and presents future prospective on the possibility of integration of high-level intelligent controller with lower-levels mechanical and kinematics features of vehicles or robotics in general concepts. | [39] |
| Learning and simulationf the human-level | Use Neuroevolution of Augmenting Topologies (NEAT) and | Artificial Neural Network | Pilots' learning curve is irregular, due to the characteristics of the problem, but presents a good positive tendency which | [47] |
| decisions involved in driving a racing car | a online videogame prototype as a test-bed environment. NEAT is a combination of ANN and Genetic Algorithm (GA) | combined to other techniques | leads them to acquire fruitful abilities in just 50 generations with a population of 120 individuals. Pilots easily learn how to turn following soft cruves, but still have big poblems ientifying and steering hard ones, which even make them crash into track limits sometimes. The paper presents individuals habing only 2 output neurons: one for turning and the other one for throttling/braking. Therefore, for this reatively young ANNs, is very dificult to acquire the high-level behaviours that have to completely change the sign of the output of the second neuron, whenevver a sharp curve is near. | |
| AV experiment (100 km) | AV following a manually driven car (trailing) | Fuzzy Logic | The authors state this paper is one of the first communications fully describing the control system and techniques required to perform an experiment with autonomous vehicles on open roads. It introduced a different control approach for controlling autonomous vehicles on urban and motorway environments. A method for online adjustment of the CACC fuzzy controller is described and implemented, coping with the most relevant disturbances and uncertain parameters, such as road slopes, passenger weight, or gear ratio. The experiment successfully proves the capability of the developed system to drive more than one hundred kilometres autonomously. A public demonstration of the described system was conducted in June 2012, comprising a 100-km route through urban and motorway environments. It was able to cope with such gaps as motorway overpasses, traffic signals, etc.The authors stated the tracking results obtained with the CACC system were very precise, with the distance error being kept to less than 1 metre. Likewise, the lateral control was able to maintain the vehicle on the path of the leader with acceptable errors for both scenarios. However, the localization system needs to be improved to allow longer GPS gaps. Also, the presence of a 900-metre-long tunnel forced the deactivation of the autonomous system while the vehicle passed through. | [40] |
| Manage and control low level vehicle actuators (steering throttle and brake) | Control schema to manage low level vehicle actuators (steering throttle and brake) based on fuzzy logic | Fuzzy Logic | The proposed automatic low-level control system has been defined, implemented and tested in a Citroen C3 testbed vehicle, whose actuators have been automated and can receive control signals from an onboard computer where the soft computing-based control system is running. The preliminary results are confirming the potential of the proposed technique. | [49] |
| Control intersection crossing (all way stops) and optimize it | Use a simulation to model and simulate the scenario. Developed a heuristic optimization algorithm for driverless vehicles at unsignalized intersections using a multi-agent system. | Optimization Heuristics | Although, the proposed research was still in its initial stages, it presented some significant time savings compared to an AWSC intersection control. It showed that by applying the proposed algorithm on only four crossing vehicles, the total delay was reduced by approximately 35 seconds, which is equivalent to a 65 percent reduction in the total intersection delay. | [42] |
| Autonomous vehicles intersections crossing optimization | Proposition of a new tool for optimizing the AVs movements through intersections - Cooperative Adaptive Cruise Control (CACC) | Optimization Heuristics | A simulation with one vehicle type and a single intersection was performed. Also, all vehicles in the simulation were assumed to have CACC system to send/receive information and follow speed advices as directed. The preliminary results are promising and encourage future studies where the authors plan to use simulations with more types of vehicles and a greater number of adjacent intersections. According to the author, the results from this research also warrant studies with regard to incorporating non-CACC vehicles into the system and studies pertaining to tackling unexpected system changes, pedestrian movements etc. | [46] |
| Path tracking cutting corners using traditional geometric algorithms | A curve safety sub-system for path tracking based on the Pure Pursuit algorithm and a dynamic look-ahead distance definition (based on vehicle current speed and lateral error). A sub-system for path tracking where an algorithm that analyzes GPS information offline classifies high curvature segments and estimates the ideal speed for each one. | Ramer-Douglas-Peucker or Ramer-Douglas algorithm | Experimental results showed improvements in comfort and safety due to the extracted geometry information and speed control, stabilizing the vehicle and minimizing the lateral error | [48] |
| Safe-optimal trajectory selection for autonomous vehicle | Use Big Data Mining approach for crash prediction and ETA. | Regression Based Models combined to other techniques | A Big Data based method and algorithm has been presented to find the safest-optimal trajectory for fully autonomous vehicles. The method proposed relies strongly on the results obtained from Big Data prediction system which predicts accidents, ETA, and clearance time. The algorithms for checking and calculating the optimal trajectory are very lightweight and straightforward. The simulations using the available data are promising. | [76] |
| How to prove the correctness of an algorithm for Vehicle Coordination | Use a distributed coordination protocol and an intersection collision avoidance (ICA) case study + Z3 Theorem prover + Satisfiability Modulo Theories (SMT) solver | Satisfiability Modulo Theories (SMT) Solver | Paper presented a formalisation of the distributed coordination problem encountered by intelligent vehicles while contending for the same physical resource. It formalised a coordination protocol and an intersection collision avoidance case study in the SMT-lib language and proved system safety using the Z3 theorem prover. The two | [43] |
| | to prove correctness and safety of a vehicular coordination problem | | main conclusions are: (1) The responsibility approach to distributed coordination is a suitable abstraction for formal reasoning on system safety. The core of this approach is that every entity is responsible for making sure that it does not enter an unsafe state with respect to any other entity. This can be contrasted with the other approaches where consensus is required between all nodes, decisions are made by a central manager, or where each pair of nodes negotiates independently, all of which seem problematic from a scalability point of view; (2) The automatic verification of collaborative vehicular applications with the help of SMT solvers is at least plausible. Some cases were found where the model could not be verified and increasing the detail and scale of the model would certainly enlarge this problem. However, there are certainly domain-specific approximations that can be made to alleviate some of these problems. | |
| Robot maneuvers too close to an obstacle increases the probability of an accident. Preventing this is crucial in dynamic environments, where the obstacles, such as other UAVs, are moving | SVM-Inspired Dynamic Safe Navigation Using Convex Hull Construction an algorithm for a fast construction of a maximum margin between sets of obstacles and its maintenance as the input data are dynamically altered | Support Vector Machine (SVM) | MMS-CH algorithm for calculating the safest path in dynamic environment was presented. It used the construction of convex hulls over the input data to eliminate data points irrelevant for the solution and to use the boundary of the hulls to search for the optimal separation margin between sets of obstacles. The tests showed the algorithm performs well in dynamic scenarios where the input data might be altered by insertion or deletion of data points. The preprocessing phase of the MMS-CH algorithm can recognize whether the change in the data set does or does not require any recalculation of the previous solution and thus prevents unnecessary computations. | [41] |
| Drivers' manoeuver classification | Motion tracking (i.e skeletal tracking) data gathered from the driver whilst driving to learn to classify the manoeuvre being performed (Kinnect) | Support Vector Machine (SVM) | Preliminary results show that skeletal tracking data can be used in a driving scenario to classify maneuvers. | [44] |
| DDT System Model Component – Machine Perception | | | | |
| Recognize leading vehicle in a convoy | Object detection using Thermal infrared classifiers and visible light classifiers | Adaptive Boosting (AdaBoost) | Thermal infrared classifiers and visible light classifiers were compared and evaluated. Both approaches perform very well. However, the accuracy of the visible light classifiers cannot be reached by thermal infra-red classifiers. But because of the good performance of the thermal infra-red classifier under all weather conditions, the performance of the thermal infra-red classifier is acceptable. | [22] |
| Prediction of advanced driver assistance systems | ML Classification techniques | Artificial Neural Network | SVM shows best ML classification performance (low errors and correlation coefficient near 1) in prognosis of the | [55] |
| (ADAS) remaining useful life (RUL) for the prognosis of ADAS' safety critical components | | combined to other techniques | ADAS systems under the given experimental assumption and Neural Networks have the worst classification performance. This work just proposes a framework for a new area of research in prognostics for automotive domain. | |
| Road environmental recognition and various object detection in real driving conditions | Single monocular camera for autonomous vehicle in real driving conditions | Artificial Neural Network combined to other techniques | Pedestrian detection algorithm with GPU were 6 times faster than CPU. Traffic sign and traffc light recognition are 2 to 3 times faster than pedestrian detection. However, when the days are dark or there is backlight, it was hard to separate the objects from background. | [27] |
| Road Terrain detection | Color feature extraction + Clustering algorithm k- mean + ANN | Artificial Neural Network combined to other techniques | Color Feature Extraction was used to classify the Road Terrain with a Neural Network (NN). 7666 images were used for classification and results were promising. | [31] |
| Road junction detection | 3D point clouds approach | Artificial Neural Network combined to other techniques | The performance of ANNs, SVMs and AdaBoost were compared for the second step of the method, and of HMMs and CRFs for the last. AdaBoost was considered the best classifier, as it managed to learn the training set without overfitting, generalizing well to the test set. On a frame-by-frame analysis, subsequent use of CRF and HMM do not seem to improve from the results obtained by AdaBoost itself. However, both methods removed a lot of the classification noise, generating an output that allows to more clearly detect the start and end of a road junction. | [30] |
| Vehicle Cyber Attack | Detection System | Artificial Neural Networks | Paper aimed to address the problem of attack detection and identification when the majority of multiple sensors was attacked in an automotive CPS. LSTM and GRU detected and identified attacks by considering sequential information with real data. It was demonstrated that the accuracy of LSTM is the highest among data-based methods (i.e., Neural Network, SVM, simple RNN, GRU and LSTM). The accuracy of LSTM followed the accuracy of GRU. Especially, LSTM and GRU had superior ability to detect coinstantaneous attacks. LSTM and GRU showed high performance in identification of Class 2, 3, 4, 5 and 6. Although calculation time of GRU is faster than that of LSTM, it is no matter to detect the attacks of the sensors on a general computer to use a CPU. | [39] |
| Turn Signal Recognition | Image Recognition and Timming (95% and 82.2% accuracy) | Artificial Neural Networks | This paper proposed the flushing light detection for preceding vehicles. The results show that the obtained classifier detects turn signals with an accuracy of 95(%). Moreover, the proposed method is capable of recognizing an appropriate frequency of flushing light with an accuracy of 82.2(%) for sequential driving data. | [18] |
| Road Sign Classification in Real-time | Use ANN (2 steps: MLP + SLP) | Artificial Neural Networks | A novel approach based on two modules was presented. The first module consists of classifying the road sign's shape using MLP. The shapes are classified in four classes: triangular, inverted triangular, circular and octagonal shapes. The accuracy of the MLP, however, is improved when using only six features values for increasing the speed of the algorithm and minimizing the memory. The second module is reserved to the identification of the contents of recognized shapes: the circular and triangular signs via a simple SLP. As for the octagonal sign and upside down triangular, they have a unique indication which are the obligation to stop and give way. A Performance Factor was introduced in order to make a subjective comparison between our proposed approach and the other methods available in the literature, which revealed that our proposed system outperforms most of the other approaches. Regarding running time, the current software implementation takes relatively a real time. | [32] |
| Spatio-temporal situation awareness | Deep Learning | Artificial Neural Networks | Given a driving video, the research aimed to model which of the surrounding vehicles are most important to the immediate driving task. Employing human-centric annotations allowed for gaining insights as to how drivers perceive different on-road objects. Although perception of surrounding agents is influenced by previous experience and driving style, we demonstrated a consistent human-centric framework for importance ranking. Extensive experiments showed a wide range of spatio-temporal cues to be essential when modeling object-level importance. Furthermore, the importance annotations proved useful when evaluating vision algorithms designed for on-road applications and autonomous driving. | [33] |
| Pedestrian detection and movement direction recognition | Deep Learning | Artificial Neural Networks | Paper presented a method to differentiate the motion of pedestrians in real life environments. Using a novel input-filtered image based on the post-processing of static recorded video frames, it could successfully distinguish three different pedestrian movement directions. Additionally, it has been proved how CNNs can impressively perform in such a task by training them with a specialized dataset. Moreover, it has been demonstrated how the results can be enhanced even further by searching for the best hyper-parameters once the CNN has been fine-tuned for the specific problem, in this case tuning the momentum and weight decay CNN parameters. Paper also presented an evaluation of the current state-of-the-art CNNs, with ResNet being the best-performing CNN for the image | [26] |
| | | | recognition problem used (94% accuracy in the validation set and 79% in the test set). | |
| Pedestrian Trajectory Prediction | Self-learning Trajectory Prediction | Artificial Neural Networks | Results show that the LSTM prediction model is superior to a constant velocity Kalman Filter for pedestrian prediction even on small datasets. Also, it was showed that the prediction model can adapt to changes in the pedestrian walking path using only a small part of the new data. By that, the size of the dataset can be kept rather small although depicting the pedestrian's movement patterns | [28] |
| Early detection of faults or malfunction | On-chip sensor diagnosis | Artificial Neural Networks | The paper discusses the suitability and feasibility of enhancing the reliability of microsensor by adding an on chip self-diagnosis capability. The approach used AI techniques and sensors with no accessible internal signals are taken as an example. Some common acceleration sensor faults are considered, and an indication is given of the manner in which these faults can be detected and isolated, either on an individual sensor basis or based on cooperative work within a sensor network. The design requirements for such self-diagnosable measument systems are set and further practical implementation issues are raised. | [55] |
| Securing connected vehicles against Denial of Service (DoS) attack | Augment message authentication with Particle Filter (PF) and it to Kalman Filter (KF). | Estimation Filters (e.g. Kalman Filter and Particle Filters) | Particle filter significantly reduces communication overhead while keeping the same detection level of spoofed messages when compared to Kalman filter in VANET applications. Stimulating different scenarios with Context adaptive beacon verification along with Kalman and particle filter on University of Massachusetts Dartmouth and State Road (Dartmouth) proved that it can detect and prevent spoofed attacks and help reducing the computational overhead. But, the Current method of securing the connected vehicle with filters leave the burden of privacy protection on VANET. The practice makes the autonomous cars the target of attack because of the number of spoofed messages missed by context adaptive beacon verification is around 11% (41 out if 46 were detected) which leaves the undetected rate too high to be replaced by conventional verification method. KF is good when road was linear and lags when the path is non-linear. PF is good for both. KF saves upto 86.5% while Particle Filter can save 85.94% computational overhead for the same scenario. Detect around 76% (24% missed) spoofed beacons with Kalman Filter and 89% (11% missed) spoofed beacons with Particle Filter. | [51] |
| Reliable and robust obstacles detection | An innovative multi-dimensional structure based on association | Estimation Filters (e.g. Kalman Filter | The presented system was able to track and fuse obstacles coming from a laser and a stereo camera. The approach has been | [25] |
| | costs originating from a classifier provides an optimal solution to the association problem with respect to the total association cost. | and Particle Filters) | compared with other state of the art algorithms, showing better results in all the considered metrics. Moreover, the system uses less computational resources and thus, fixed the platform, may work at higher frame rate compared to other solutions, making it more appropriate for automotive applications. The system has demonstrated a correct reconstruction of the dynamic world surrounding the vehicle, proving to be able to help the driver in the assessment of critical situations. In particular, the developed algorithm provides a stable, robust and reliable detection, classification and tracking of the multiple targets coming from different sources. Moreover, the proposed approaches were seen to outperform the state-of-the-art approaches on a public dataset. A fault tolerant and reliable system requires sensors redundancy and complementarity. Common approaches rely on object level fusion. It has been introduced a medium-level fusion which take advantage from both the approaches. The fusion is performed at object level but preserving the low-level information; in this way it is guaranteed a real-time processing exploiting all available information. | |
| Cyber Attack in V2X | Fuzzy Detector | Fuzzy Logic | To address security issues of a system of connected vehicles (CVs), a fuzzy detector was also introduced that detects possibility of a cyber threat and takes proper actions in response to the specific attack. Results show the designed system can detect any adversary access to the system and can prevent subsequent crashes by adjusting the safe following distance. | [54] |
| Road and Obstacle Detection | Sensor Fusion | Fuzzy Logic | A high-level data-fusion strategy has been devised, which is based on the identification and representation of the descriptive and procedural knowledge required. Such a strategy yields better recognition results by merging the various hypotheses generated by the separate channels and solving possible conflicts through a fuzzy representation of knowledge when compared to a benchmark. In addition, the data-fusion system has performed a more accurate segmentation process by using goal-driven low-level procedures, according to which the image regions have been assigned relative fuzzy memberships to the object to be recognized. | [24] |
| Enhance Image Understanding | Develop generic technology that will enable the construction of complete, robust, high performance image understanding systems | General AI/ML | This paper provided an overview of the technical and program management plans being used in evolving the proposed technology, but no results were presented. | [19] |
| | to support a wide range of DoD applications | | | |
| Learn from Demonstration | Use a Semi-automated mine robot | Hidden Markov Based Models (e.g. Continuous Hidden Markov Model-CHMM and Discrete Hidden Markov Model-DHMM) combined to other techniques | In this paper, three methods were compared based on three trajectories in the low noise and noisy environments. The GMM based method had the best performance in a low noise environment. In practice, there's always unexpected noise around a robot, implying the GMM based method was not practical for real environments. The CHMM based method was suitable for turning trajectories, while The DHMM based method was more robust for straight trajectories. | [45] |
| Vehicle Detection and Counting | Hidden Markov Model + Viterbi algorithm + Adaboost trained Haar-like feature detector Approach | Hidden Markov Based Models (e.g. Continuous Hidden Markov Model-CHMM and Discrete Hidden Markov Model-DHMM) combined to other techniques | The proposed method has been shown to give significantly better vehicle volume counts than both multiple targets moving object tracking and VDL on a dataset of over 88 hours of video. On this testing set, the proposed method achieved a median 5-minute-bin error of 0.0686 for this counting task while the multiple target motion tracking and VDL implementations had median errors of 0.0957 and 0.2290 respectively. The proposed method was also more reliable having fewer and less severe occurrences of 5-minute-bin errors throughout the testing set. | [36] |
| Safety-warning and driver-assistance system and an automatic pilot for rural and urban traffic environments | Adaptive randomized HT (RHT) for robust and accurate detection of lane markings without manual initialization or a priori information under different road environments | Hough Transformation | In this paper, a prototype was demonstrated, and tasks of lane detection detailed. Preliminary experimental results in different road scene and a comparison with other methods have proven the validity of the proposed method | [21] |
| Speed bump detection | Use Camera and image recognition | Novel Image Recognition Technique | The average performance of the system considering only speed bump with proper marking is 85%. | [34] |
| Obstacle clusteering and tracking | Lidar + split-and-merge algorithm | Ramer-Douglas-Peucker or Ramer-Douglas algorithm | Paper presents a robust platform for implementing a perception system for ground vehicles using a LIDAR sensor and two cameras has been designed. Tests were performed using this platform and different implementations, and the results were checked with the real-world scenes, demonstrating the technique validity. | [23] |
| Predicts the CPU usage patterns of software tasks running on a self-driving car | Methods for learning the patterns of tasks' CPU utilizations in given driving contexts | Regression Based Models | A feature vector was designed to represent the internal and external states of a self-driving car and five regression methods were used to predict the CPU usage patterns of four software tasks running on a self-driving car. Through testing with the actual driving data, the results showed a regression method could be used to predict a software task's dynamic | [53] |
| | | | CPU utilization. | |
| Providing road safety to connected drivers and connected autonomous vehicles | Observing the Doppler profile | Support Vector Machine (SVM) | The paper presented a collision and driving scenario classification system based on the Doppler profile which could potentially decouple the safety benefits of V2V communications from relying on SM content. The Doppler profile in V2V networks showed rich data about the vehicles and their environments and could be exploited to potentially provide a reliable collision avoidance service directly from the radio front-end. No experimental results were presented. | [20] |
| Road Detection | Road area detection method using a support vector machine (SVM) and histogram of oriented gradient (HOG) features and 3D lidar | Support Vector Machine (SVM) combined to other techniques | Classifier to differentiate road areas from other areas using a 3D Lidar with machine larning techniques. Range data of lidar changes in the layer direction but not in the rotational direction. HOG features of the reaos concentrate in the same bin. In contract the features of the non-road areas are distributed among several bins representing various directions. Found differences between the histograms for the roas planes and the other areas. In real world data, same tendences of HO features. Error rate of 8.51% using SVM. Area up to 10m ahead of the vehicle can be identified correctly. | [75] |
| Traffic Light Detection | Two-stage preprocessing using Principal Component Analysis (PCA) followed by SVM | Support Vector Machine (SVM) combined to other techniques | Paper presents a system that can detect multiple types of green and red traffic lights accurately and reliably. Color extraction and blob detection were applied to locate the candidates with proper optimization. A classification and validation method using PCANet was then used for frame-by-frame detection. Multiobject tracking method and forecasting technique were succesfully employed to improve accuracy and produced stable results. | [35] |
| Pedestrian Detection | Use High-Definition 3D Range Data (from a LIDAR) | Support Vector Machine (SVM) combined to other techniques | An exhaustive analysis of the performance of three different machine learning algorithms have been carried out: k-Nearest Neighbours (kNN), Naïve Bayes classifier (NBC), and Support Vector Machine (SVM). Each algorithm was trained with a training set comprising tool 277 pedestrians and 1654 no pedestrian samples and different kernel functions: kNN with Euclidean and Mahalanobis distances, NBC with Gauss and KSF functions and SVM with linear and quadratic functions. LOOCV and ROC analysis were used to detect the best algorithm to be used for pedestrian detection. The proposed algorithm has been tested in real traffic scenarios with 16 samples of pedestrians and 469 samples of non-pedestrians. The results obtained were used to validate theoretical results. An overfitting problem in the SVM with | [27] |
| | | | quadratic kernel was found. Finally, SVM with linear function was selected since it offered the best results. A comparison of the proposed method with five other works that also use High Definition 3D LIDAR to carry-out the pedestrian detection, comparing the AUC and Fscore metrics. Conclusions are the proposed method obtains better performance results in every case. Pedestrian detection has traditionally been performed using machine vision and cameras, but these techniques are affected by changing lighting conditions. 3D LIDAR technology provides more accurate data (more than 1 million points per revolution), which can be successfully used to detect pedestrians in any kind of lighting conditions | |
| How to "automate" manual annotation for images to train visual perception for AVs | Training visual models using photo-realistic computer graphics | Support Vector Machine (SVM) combined to other techniques | The experiments showed that virtual-world data is effective for training vision-based pedestrian detectors which can be adapted to operate in real scenarios. The different adaptation procedures have shown to provide adapted detectors that improve those trained only on virtual data, as well as those trained using only the real-world data available for the adaptation (which constrained to save a ∼90% annotation effort along the experiments). | [37] |